\documentclass[fleqn,usenatbib]{mnras}
\usepackage[T1]{fontenc}

\DeclareRobustCommand{\VAN}[3]{#2}
\let\VANthebibliography\thebibliography
\def\thebibliography{\DeclareRobustCommand{\VAN}[3]{##3}\VANthebibliography}

\usepackage{amsmath}
\usepackage{booktabs}
\usepackage{textcomp} % Prevents warnings about not defining \perthousand and \micro.
\usepackage{gensymb}
\usepackage[utf8]{inputenc}
\usepackage{multirow}
\usepackage{rotating}
\usepackage{savesym}
\savesymbol{tablenum} % Workaround for conflict of tablenum definition in aastex63.cls.
\usepackage{siunitx}
\restoresymbol{SIX}{tablenum}
\usepackage{tabularx}
\usepackage{tikz}
\usepackage{ulem}
\usepackage{xifthen}
\usepackage{xspace}

\graphicspath{{./}{figures/}}

\usepackage{graphicx}	% Including figure files
\usepackage{amsmath}	% Advanced maths commands
\usepackage{amssymb}	% Extra maths symbols

\newcommand{\Planck}{\textit{Planck}\xspace}
\newcommand{\Suzaku}{\textit{Suzaku}\xspace}
\newcommand{\AIC}{\ensuremath{\mathrm{AIC}}}
\newcommand{\fil}{\ensuremath{\mathrm{fil}}}

\newcommand{\modelname}[2]{\newcommand{#1}[1][]{{{\ifthenelse{\isempty{##1}}{`}{}}#2\ifthenelse{\isempty{##1}}{'}{}}\xspace}}
\modelname{\modnobridge}{Ellip--$\beta$, no bridge}
\modelname{\modthreebeta}{3$\times$Ellip--$\beta$}
\modelname{\modcircmesa}{Circ--$\beta$+mesa}
\modelname{\modmesa}{Ellip--$\beta$+mesa}

\DeclareSIUnit{\parsec}{pc}
\DeclareSIUnit{\arcminute}{arcmin}
\DeclareSIUnit{\year}{yr}
\newcommand{\msun}{\ensuremath{\mathrm{M}_{\odot}}}
\newenvironment{enumerate*}{\begin{enumerate}%
        \setlength{\itemsep}{0pt}%
        \setlength{\parskip}{0pt}%
}{\end{enumerate}}

\newcommand{\const}[3]{\newcommand{#1}[1][]{ {\ensuremath{#2\ifthenelse{\isempty{##1}}{\,#3}{}}}}}

\const{\sensitivityturnover}{\gtrsim 6'}{}
\const{\ymapnoise}{3.0\times10^{-6}}{\mathrm{arcmin}}
\const{\ymapdeprojnoise}{3.5\times10^{-6}}{\mathrm{arcmin}}
\const{\dustdeprojerrincrease}{{\sim}50}{\%}
\const{\clustervsclusterbridge}{> 5\sigma}{}
\const{\bridgedensity}{(0.88\pm0.24)\times10^{-4}}{\si{\per\centi\meter\cubed}}
\const{\Y}{(6.6\pm1.4)\times10^{-3}}{\mathrm{arcmin}^{2}}
\const{\Ympc}{(4.9\pm1.0)\times10^{-5}}{\si{\mega\parsec^2}}
\const{\bridgegasmass}{(5.3\pm1.1)\times10^{13}}{\msun}
\const{\bridgemass}{(3.3\pm0.7)\times10^{14}}{\msun}
\const{\bridgemasspercent}{7.8\%}{}
\const{\bridgemasspercentapprox}{8\%}{}
\const{\bridgetotaly}{(2.8\pm0.3)\times10^{-5}}{}
\const{\bridgecirctotalyapprox}{2.5\times10^{-5}}{}
\const{\mesalengthmpc}{(2.2\pm0.3)}{\si{\mega\parsec}}
\const{\mesalengthmpcapprox}{2.2}{\si{\mega\parsec}}
\const{\mesawidthmpc}{(1.9\pm0.2)}{\si{\mega\parsec}}
\const{\mesawidthmpcapprox}{1.9}{\si{\mega\parsec}}
\const{\mesaheightmpc}{(12.1\pm3.9)}{\si{\mega\parsec}}
\const{\mesaheightmpcapprox}{12.1}{\si{\mega\parsec}}
\const{\mesaamplitude}{1.10^{+0.17}_{-0.18}\times10^{-5}}{}
\const{\mesaw}{43.4}{}
\const{\mesap}{3.1\times10^{-8}}{}
\const{\mesapref}{5.5}{\sigma}
\const{\mesaprefapprox}{5}{\sigma}
\const{\mesaprefplanck}{3.6}{\sigma}
\const{\mesaprefplanckapprox}{3.5}{\sigma}
\const{\threebetaw}{33.2}{}
\const{\threebetap}{3.4\times10^{-6}}{}
\const{\threebetapref}{4.6}{\sigma}
\const{\circvsellipw}{10.4}{}
\const{\circvsellipaicw}{2.4}{}
\const{\circvsellipsig}{2.1}{\sigma}
\const{\circvsellipp}{0.034}{}
\const{\circvsellipsigddp}{1.0}{\sigma}
\const{\athreemassf}{(8.7\pm1.0)\times10^{14}}{\msun}
\const{\afourmassf}{(10.2\pm1.3)\times10^{14}}{\msun}
\const{\athreemasst}{(18.1\pm2.2)\times10^{14}}{\msun}
\const{\afourmasst}{(21.3\pm2.7)\times10^{14}}{\msun}
\const{\athreeradiusf}{(1.45\pm0.21)}{\si{\mega\parsec}}
\const{\afourradiusf}{(1.53\pm0.19)}{\si{\mega\parsec}}
\const{\suzakugasmass}{4.8\times10^{13}}{\msun}
\const{\suzakutotalmass}{3.0\times10^{14}}{\msun}
\const{\toymodelnaifangle}{9.0^{\circ}}{}
\const{\toymodelangle}{16.6^{+5.5}_{-3.8}}{\mathrm{deg}}
\const{\toymodelangleapprox}{17}{^{\circ}}
\const{\toymodelseparation}{11.1^{+3.2}_{-2.6}}{\si{\mega\parsec}}
\const{\toymodelseparationapprox}{11}{\si{\mega\parsec}}
\const{\toymodelvpec}{542^{-9}_{+18}}{\si{\kilo\meter\per\second}}
\const{\toymodelinitseparation}{12.3^{+4.0}_{-3.1}}{\si{\mega\parsec}}
\const{\toymodelinittime}{4.4^{+3.0}_{-1.8}}{\si{\giga\year}}
\const{\toymodelageuniverse}{15.7^{+8.9}_{-5.9}}{\si{\giga\year}}
\const{\toymodelellipangle}{14.3^{+4.6}_{-3.2}}{\mathrm{deg}}
\const{\athreer}{0.93}{}
\const{\afourr}{0.82}{}
\const{\toymodelellipathreer}{0.53}{}
\const{\toymodelellipafourr}{0.33}{}
\const{\afilplanckonly}{(0.89\pm0.19)\times10^{-5}}{}
\const{\afilplanckonlysig}{4.7}{\sigma}
\const{\afilplanckonlyfixpossig}{5.8}{\sigma}
\const{\afilplanckonlywhitenoisesig}{9.7}{\sigma}
\const{\clusterseparation}{3.2}{\si{\mega\parsec}}

\title[A view of A399--401 with ACT and MUSTANG-2]{A high-resolution view of the filament of gas between Abell~399 and Abell~401 from the Atacama Cosmology Telescope and MUSTANG-2}

\author[A.~D.~Hincks~et~al.]{%
Adam~D.~Hincks,$^{1}$\thanks{E-mail: adam.hincks@utoronto.ca}
Federico~Radiconi,$^{2}$
Charles~Romero,$^{3,4}$
Mathew S.~Madhavacheril,$^{5,6}$
\newauthor
Tony~Mroczkowski,$^{7}$
Jason~E.~Austermann,$^{8}$
Eleonora~Barbavara,$^{2}$
Nicholas~Battaglia,$^{9}$
\newauthor
Elia~Battistelli,$^{2}$
J.~Richard~Bond,$^{10}$
Erminia~Calabrese,$^{11}$
Paolo~de Bernardis,$^{2}$
Mark~J.~Devlin,$^{4}$
\newauthor
Simon R.~Dicker,$^{4}$
Shannon~M.~Duff,$^{8}$
Adriaan~J.~Duivenvoorden,$^{12}$
Jo~Dunkley,$^{12,13}$
\newauthor
Rolando~D\"unner,$^{14}$
Patricio~A.~Gallardo,$^{15}$
Federica~Govoni,$^{16}$
J.~Colin~Hill,$^{17,18}$
Matt~Hilton,$^{19}$
\newauthor
Johannes~Hubmayr,$^{8}$
John~P.~Hughes,$^{20}$
Luca~Lamagna,$^{2}$
Martine~Lokken,$^{1,10,21}$
Silvia~Masi,$^{2}$
\newauthor
Brian~S.~Mason,$^{22}$
Jeff~McMahon,$^{23,24,25,26}$
Kavilan~Moodley,$^{19,27}$
Matteo~Murgia,$^{16}$
Sigurd~Naess,$^{18}$
\newauthor
Lyman~Page,$^{12}$
Francesco~Piacentini,$^{2}$
Maria~Salatino,$^{28,29}$
Craig~L.~Sarazin,$^{30}$
\newauthor
Alessandro~Schillaci,$^{31}$
Jonathan~L.~Sievers,$^{32,33,34}$
Crist\'obal~Sif\'on,$^{35}$
Suzanne~Staggs,$^{12}$
\newauthor
Joel~N.~Ullom,$^{8}$
Valentina~Vacca,$^{16}$
Alexander~Van~Engelen,$^{36}$
Michael~R.~Vissers,$^{8}$
\newauthor
Edward~J.~Wollack$^{37}$~and
Zhilei~Xu$^{4,38}$
\newauthor
\newline
\textit{\normalsize Affiliations are listed at the end of the paper.}%
}%
\date{Accepted XXX. Received YYY; in original form ZZZ}

\pubyear{2021}

\begin{document}
\label{firstpage}
\pagerange{\pageref{firstpage}--\pageref{lastpage}}
\maketitle

\begin{abstract}
    We report a significant detection of the hot intergalactic medium in the filamentary bridge connecting the galaxy clusters Abell~399 and Abell~401. This result is enabled by a low-noise, high-resolution map of the thermal Sunyaev-Zeldovich signal from the Atacama Cosmology Telescope (ACT) and \Planck satellite. The ACT data provide the $1.65'$ resolution that allows us to clearly separate the profiles of the clusters, whose centres are separated by $37'$, from the gas associated with the filament. A model that fits for only the two clusters is ruled out compared to one that includes a bridge component at $> \mesaprefapprox$. Using a gas temperature determined from \Suzaku X-ray data, we infer a total mass of $\bridgemass$ associated with the filament, comprising about $\bridgemasspercentapprox$ of the entire Abell~399--Abell~401 system. We fit two phenomenological models to the filamentary structure; the favoured model has a width transverse to the axis joining the clusters of ${\sim}\mesawidthmpcapprox$. When combined with the \Suzaku data, we find a gas density of $\bridgedensity$, considerably lower than previously reported. We show that this can be fully explained by a geometry in which the axis joining Abell~399 and Abell~401 has a large component along the line of sight, such that the distance between the clusters is significantly greater than the $\clusterseparation$ projected separation on the plane of the sky. Finally, we present initial results from higher resolution ($12.7\arcsec$ effective) imaging of the bridge with the MUSTANG-2 receiver on the Green Bank Telescope. 
\end{abstract}


\section{Introduction}
\label{sec:introduction}

A significant discrepancy between the total quantity of baryonic matter readily observable in the local Universe and that measured at high redshift in the Lyman-alpha forest or predicted by big bang nucleosynthesis has long been noted. Early estimates by \citet{fukugita/hogan/peebles:1998} (see also \citealt{fukugita/peebles:2004}) indicated that these `missing baryons' could be in plasma outside of galaxy clusters, a hypothesis that found robust support in hydrodynamical simulations \citep{cen/ostriker:1999,cen/ostriker:2006}. According to this theory, a significant fraction must reside in the diffuse (${\sim}10{-}100$ times the mean baryon density), ${\sim}10^5{-}10^7\,\si{\kelvin}$ gas known as the warm-hot intergalactic medium (WHIM), with ${\sim}30\%$ of the baryons in the low-redshift universe found in the outskirts of galaxy clusters and along the filaments of the dark matter web connecting them (\citealt{tuominen/etal:2021}; see also \citealt{shull/smith/danforth:2012,martizzi/etal:2019,galarraga-espinosa/etal:2021}). Measurements of the frequency dispersion of localized fast radio bursts, which is determined by the total quantity of plasma along the line of sight, have recently provided results consistent with the predictions for the total baryon density \citep{macquart/etal:2020}, and multiple types of observation now support the expectations for where the missing baryons are located. Stacking- and power spectrum-based studies with Sunyaev-Zeldovich (SZ) effect (introduced below) are one line of evidence. Measurements of the kinematic SZ effect, which is proportional to the average momentum of ionized gas, have revealed a large quantity of baryons beyond the virial radii of clusters, as predicted by theory \citep{planck/etal:2016,tanimura/etal:2021,schaan/etal:2021,kusiak/etal:2021}. The signal from the thermal SZ effect, which depends on the gas pressure, has been detected between stacked pairs of galaxies \citep{degraaff/etal:2019,tanimura/etal:2019,tanimura/etal:2020a}, indicative of baryons in intercluster filaments. A stacking approach has also recently been successfully applied to \textit{ROSAT} X-ray data \citep{tanimura/etal:2020b}. A final technique is to search for absorption lines in the spectra of quasars caused by intervening WHIM, and recent possible detections have been made with  \ion{O}{vi}, \ion{O}{vii}, \ion{O}{viii} and \ion{H}{i} absorption \citep[e.g.,][]{tejos/etal:2016,nicastro/etal:2018,pessa/etal:2018,Nevalainen2019,bouma/richter/wendt:2021}.

Still, directly imaging the distribution of the WHIM---that is, without stacking---remains challenging. Its low density and intermediate temperatures mean that its X-ray emission is faint and therefore difficult to detect \citep{bregman:2007}, and the low number of relativistic electrons and the weakness of the expected magnetic fields makes radio observations of synchrotron radiation challenging \citep{vazza/etal:2015}.

\begin{table*}
    \caption{Basic properties of the A399--401 system. Cluster coordinates (in J2000) and redshifts are from the NASA/IPAC Extragalactic Database (NED).}
    \label{tab:basic_properties}
    \centering
    \begin{tabular}{cccccc}%
        \hline\hline
             & RA & Dec & $z$ & \multicolumn{2}{c}{Separation in Plane of Sky} \\
             & & & & (arcmin) & (Mpc) \\
        \hline
        A399 & 02h57m56.4s & +13$^{\circ}$00$'$59$''$ & 0.071806 & \multirow{2}{*}{36.9} & \multirow{2}{*}{\clusterseparation[1]} \\
        A401 & 02h58m57.5s & +13$^{\circ}$34$'$46$''$ & 0.073664 & & \\
        \hline
    \end{tabular}%
\end{table*}

The cluster pair formed by Abell~399 and Abell~401 (hereafter A399 and A401, respectively, or A399--401 for the system as a whole) is a rare system where direct imaging is currently possible. The clusters are separated by $37'$, corresponding to a proper separation of $\clusterseparation$ in the plane of the sky (see Table~\ref{tab:basic_properties}). The first evidence for a bridge of plasma between the clusters came from a measurement of excess X-ray emission in the intercluster region using the Advanced Satellite for Cosmology and Astrophysics (ASCA; \citealt{fujita/etal:1996,markevitch/etal:1998}). \citet{fabian/peres/white:1997} observed A399 with the \textit{ROSAT} High Resolution Imager and found a structure extending from the cluster towards A401. The presence of hot gas, ${\sim}6$--$7\,\si{\kilo\electronvolt}$ (${\sim}7{-}8\times10^{7}\,\si{\kelvin}$), between A399 and A401 has since been confirmed by observations with the X-ray Multi-Mirror Mission (\textit{XMM-Newton}; \citealt{sakelliou/ponman:2004}; c.f., \citealt{bourdin/mazzotta:2008}) and \Suzaku \citep{fujita/etal:2008,akamatsu/etal:2017}. This gas is thus hotter (and denser; see below) than the majority of the WHIM \citep{degraaff/etal:2019}. \citet{akamatsu/etal:2017} report evidence of an equatorial shock, that is, parallel to the axis joining the clusters. Furthermore, each of the clusters hosts a radio halo \citep{murgia/etal:2010} and recently \citet{govoni/etal:2019} detected a `ridge' of radio emission between them using 140\,MHz Low Frequency Array (LOFAR) observations. \citet{nunhokee/etal:2021}, however, did not detect this ridge in their 346\,MHz Westerbork observations, which they conclude must be due to a steep spectral index ($\alpha < -1.5$ at 2$\sigma$).\footnote{We adopt the convention $S \propto (\nu/\nu_0)^{\alpha}$ for flux density $S$, frequency $\nu$, reference frequency $\nu_0$ and spectral index $\alpha$.}

The consensus is that A399 and A401 have not interacted in the past -- contrary to earlier speculation \citep{fabian/peres/white:1997} -- and are likely in a pre-merger phase. The X-ray data show a smooth temperature profile in the bridge without any evidence for large shocks or similar disruptions that would indicate prior interaction \citep{fujita/etal:1996,markevitch/etal:1998,sakelliou/ponman:2004}, and while the morphologies of A399 and A401 exhibit irregularities, they are consistent with mergers happening in each cluster independently \citep{bourdin/mazzotta:2008} which could reasonably account for why they have radio halos \citep{murgia/etal:2010}. Moreover, tidal forces do not seem sufficient to cause the high temperatures in the bridge, but which could be due to compression of the WHIM by the clusters' motion towards each other (\citealt{sakelliou/ponman:2004}; see also \citealt{akahori/yoshikawa:2008,akamatsu/etal:2017}). The radio emission along the bridge might come from low-level shocks in gas falling toward the filament that accelerate electrons in magnetic fields between the clusters \citep{govoni/etal:2019}. Another possibility is that pre-existing relativistic particles and magnetic fields are re-energized by the dissipation of turbulence along the filament \citep{brunetti/vazza:2020}; \citet{nunhokee/etal:2021} argue that the steep radio spectral index they infer favours this scenario. Finally, \citet{fujita/etal:2008} found that the metallicity in the bridge region, ${\sim}0.2\,{\rm Z}_{\odot}$, is essentially the same as in the clusters, and speculate that the gas in which the A399--401 system formed had already been seeded by superwinds blowing metals out of earlier galaxies.

Another way to probe the gas in A399--401 is with the Sunyaev-Zeldovich (SZ) effect, or the inverse Compton scattering of the cosmic microwave background (CMB) radiation by hot electrons (\citealt{zeldovich/sunyaev:1969}, \citealt{sunyaev/zeldovich:1972}; see \citealt{mroczkowski/etal:2019} for a recent review on the application of SZ observations for astrophysical studies). The SZ signal separates into two main components. The kinematic SZ effect is due to the bulk momentum of the gas, and is therefore proportional to the peculiar velocities of galaxy clusters and other large scale structures. It follows the same blackbody spectrum as the CMB. The thermal SZ effect, on the other hand, is a frequency-dependent distortion of the CMB due to the pressure of the gas; it is this observable that we exploit in this paper, and henceforth all references to the SZ implicitly denote the thermal effect. Its amplitude is proportional to the Compton $y$-parameter:
\begin{equation}
    y = \frac{\sigma_\textsc{t}}{m_{\rm e}c^2} \int P_{\rm e} (r) dr = \frac{\sigma_\textsc{t}}{m_{\rm e}c^2} \int n_{\rm e} (r) k_\textsc{b} T_{\rm e}(r) dr
    \label{eq:tsz}
\end{equation}
where $\sigma_\textsc{t}$ is the Thomson cross section, $m_{\rm e}$ is the electron mass, $c$ the speed of the light, $k_\textsc{b}$ the Boltzmann constant, $r$ denotes the distance along the line of sight, $P_{\mathrm{e}}$ is the gas pressure and $T_{\rm e}(r)$ and $n_{\mathrm{e}}(r)$ are the electron temperature and density, respectively. The SZ distortion has a frequency dependence which, in the non-relativistic limit, modifies the thermodynamic temperature of the CMB as
\begin{equation}
    \Delta T = T_{\textsc{cmb}} \, y \, \left[x\coth(x/2) - 4\right]
    \label{eq:sz_spec}
\end{equation}
where $T_\mathrm{CMB} = (2.7260 \pm 0.0013)\,\si{\kelvin}$ is the mean temperature of the CMB \citep{fixen:2009} and $x \equiv h\nu / k_\textsc{b}T_{\mathrm{CMB}} = \nu / (56.79\,\si{\giga\hertz})$, where $h$ is the Planck constant. At frequencies below the `null' at 217\,GHz, the CMB surface brightness is reduced. Given the linear dependence of $\Delta T$ on the gas density $n_{\mathrm{e}}$, diffuse low density signals are, in principle, easier to measure with the SZ effect with respect to X-Ray emission, whose brightness goes as $L_{\textsc{x}} \propto n_{\mathrm{e}}^2 $.

The first spatially resolved SZ measurements of A399 and A401 \citep{udomprasert/etal:2004} lacked the sensitivity and frequency coverage to detect inter-cluster plasma, partly due to the noise introduced by primary CMB anisotropies. However, the \citet{planck/etal:2013} reported an excess SZ signal in the region between A399 and A401. They interpreted this as the first SZ detection of filamentary gas, reporting a density of $(3.7\pm0.2)\times10^{-4}\,\si{\per\centi\meter\cubed}$ in a joint fit with X-ray data. \cite{bonjean/etal:2018} used \Planck's 2015 maps to confirm this result, finding a density of $(4.3\pm0.7)\times10^{-4}\,\si{\per\centi\meter\cubed}$. They also used optical and infrared data to show that the galaxies in the bridge, like those of the clusters, belong to an old and red population, consistent with the bridge being a primordial cosmic filament. The \citet{planck/etal:2013} notes that the relatively low resolution of their Compton-$y$ maps ($10'$ in the component-separated maximum internal linear component analysis (MILCA)  maps) is related to their ability to recover larger angular scale signals compared to X-ray, thereby providing  sensitivity to the faint SZ signal in the cluster bridge.
At the same time, it means that the bridge is only ${\sim}3$ beams in length, and the authors note that higher resolution SZ data would help prove that the gas in the bridge is not simply part of the clusters' outskirts \citep[see also][]{bonjean/etal:2018}. Separating the clusters' contribution from that of the WHIM in the filament would provide compelling support for the prediction that missing baryons reside in the intercluster web.

In this paper, we use data from the Atacama Cosmology Telescope (ACT) to provide the higher resolution ($1.65\arcmin$) required to address this issue and to provide an improved measurement of the SZ amplitude of the bridge. Additionally, we provide initial results from even higher resolution observations (${\sim}12.7\arcsec$ after smoothing) of A399--401 with the MUSTANG-2 instrument \citep{dicker/etal:2014} on the 100-meter Green Bank Telescope (GBT) to probe small-scale fluctuations within the bridge. We describe our data and maps in Sec.~\ref{sec:data}. In Sec.~\ref{sec:results} we fit the ACT $y$-map with a series of models that both include and exclude a bridge component in order to determine whether the data favour its inclusion. Then, in Sec.~\ref{sec:discussion} we show that our fits constitute a detection of excess gas in the bridge, estimate the total mass of the bridge, provide a measurement of its density by combining with previous X-ray measurements, and show that our results are consistent with a geometry in which the axis joining A399 and A401 is largely out of the plane of the sky. We follow this in Sec.~\ref{sec:features} with a brief discussion of small-scale features in the bridge, including an interpretation of the MUSTANG-2 results. We conclude in Sec.~\ref{sec:conclusion}.

Throughout, we assume a $\Lambda$CDM flat universe with $H_0 = 67.6\,\si{\kilo\meter\per\second\per\mega\parsec}$, $\Omega_m = 0.31$ and $\Omega_{\Lambda} = 0.69$ \citep{aiola/etal:2020}. For the redshift of the filament between A399 and A401, we adopt $z = 0.072735$, the mean of the clusters' redshifts (Table~\ref{tab:basic_properties}). Distances are proper rather than comoving.

\begin{figure*}
    \centering
    {\includegraphics[clip, trim=0.0cm 0.32cm 1.96cm 0.7cm, height=7.9cm]{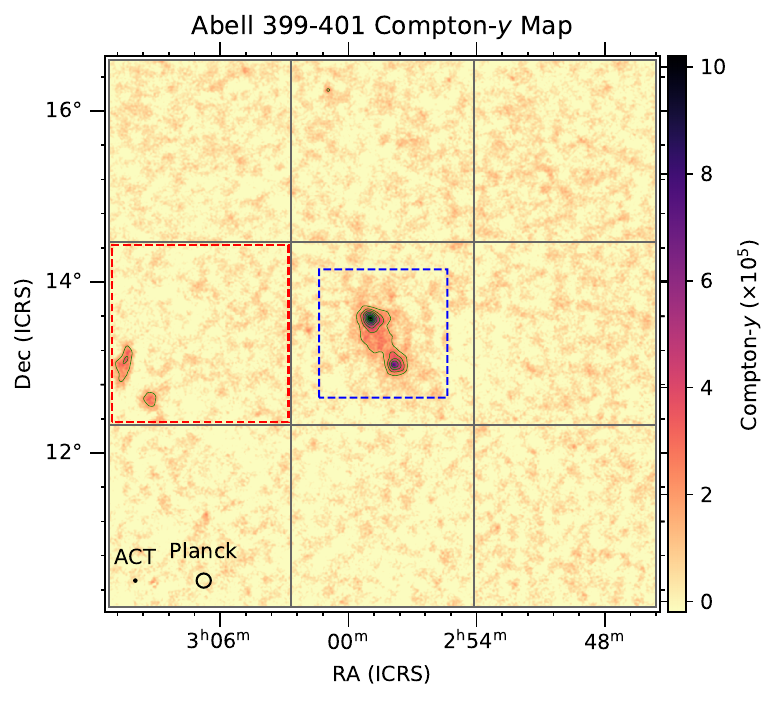}}
    {\includegraphics[clip, trim=0.6cm 0.32cm 0.20cm 0.7cm, height=7.9cm]{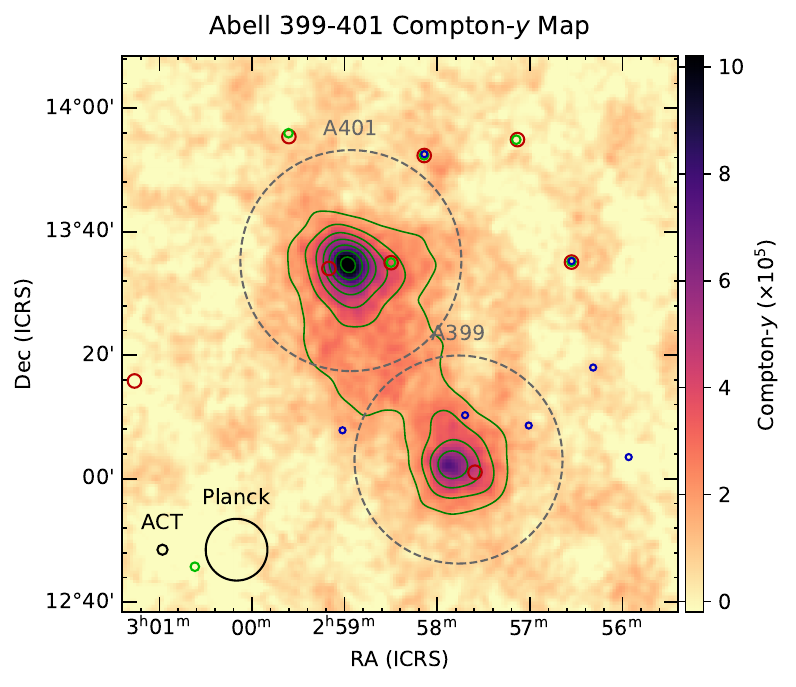}}
    \caption{{\bf Left:} Compton-$y$ map of A401 (north-east) and A399 (south-west) made with ACT and \Planck data (see text for details). Contour levels are at 3, 5, 7, 9, 11, 13, and 15$\sigma$.  Of the eight perimeter regions (grey boxes), the region to the east (marked with a dashed red box) is excluded from the covariance analysis due to dust contamination.  The dashed blue region denotes the zoom-in shown to the right.
    {\bf Right:} The panel shows a zoom-in on the dashed blue region in the left panel, with the clusters denoted by dashed circles corresponding to their measured $R_{500\mathrm{c}}$ (the radius inside of which the average density of the cluster is $500$ times the critical density of the Universe; see Table~\ref{tab:phys_properties}).
    The solid black circles in the lower left of each panel show the $1.65\arcmin$ effective beam size of our map, driven by ACT's high resolution, as well as the $10\arcmin$ resolution of the \Planck-only MILCA $y$-map (see Sec.~\ref{ssec:stats}). 
    Compact sources detected in the ACT maps are represented by coloured circles, where the sizes reflect the FWHM of the beam at that frequency and the colours -- red, blue, and green -- indicate the 98, 150, and 224~GHz data.
    The colour scale is the same in both panels.}
    \label{fig:act_y_map}
\end{figure*}

\section{Data}
\subsection{ACT Data and Compton-\texorpdfstring{$y$}{y} Map}
\label{ssec:act_data}
\label{sec:data}

ACT is a 6-metre, off-axis Gregorian telescope that began observations in 2007. It has had three generations of receivers employing transition-edge sensor (TES) bolometers: the Millimeter Bolometric Array Camera (MBAC) observed at 150, 220 and 280\,GHz from 2008 to 2010 \citep{swetz/etal:2011}, ACTPol added polarization sensitivity and observed at 98 and 150\,GHz from 2013 to 2016 \citep{thornton/etal:2016}, and Advanced ACTPol \citep{henderson/etal:2016}, whose initial deployment overlapped with ACTPol, has been observing since 2016 at 98, 150 and 220\,GHz, with 30 and 40 GHz added in 2020.\footnote{All band centres listed here are approximate.} In this paper, we use ACT maps at 98, 150 and 220\,GHz made using data from all three receivers from 2008 to 2019; the main difference from the maps of Data Release~5 (DR5), described in \citet{naess/etal:2020}, is the inclusion of 2019 data. The white noise rms of these maps in the region of A399--401 is $19\,\si{\micro\kelvin}$-arcmin (98\,GHz), $16\,\si{\micro\kelvin}$-arcmin (150\,GHz) and $66\,\si{\micro\kelvin}$-arcmin (220\,GHz).\footnote{Throughout, map temperatures are with reference to the CMB blackbody spectrum (sometimes denoted $K_{\mathrm{CMB}}$) rather than Rayleigh-Jeans temperatures.} Since atmospheric contamination increases the noise at larger angular scales, we include \Planck maps in our construction of Compton-$y$ maps to improve sensitivity on these larger scales ($\sensitivityturnover$ for the map used in this paper, described below). This also allows us to take advantage of \Planck's 353 and 545\,GHz channels for better removal of galactic dust and cosmic infrared background (CIB) emission.

\begin{figure}
    \centering
    {\includegraphics[clip, trim=0.0cm 0.32cm 0.2cm 0.7cm, height=6.9cm]{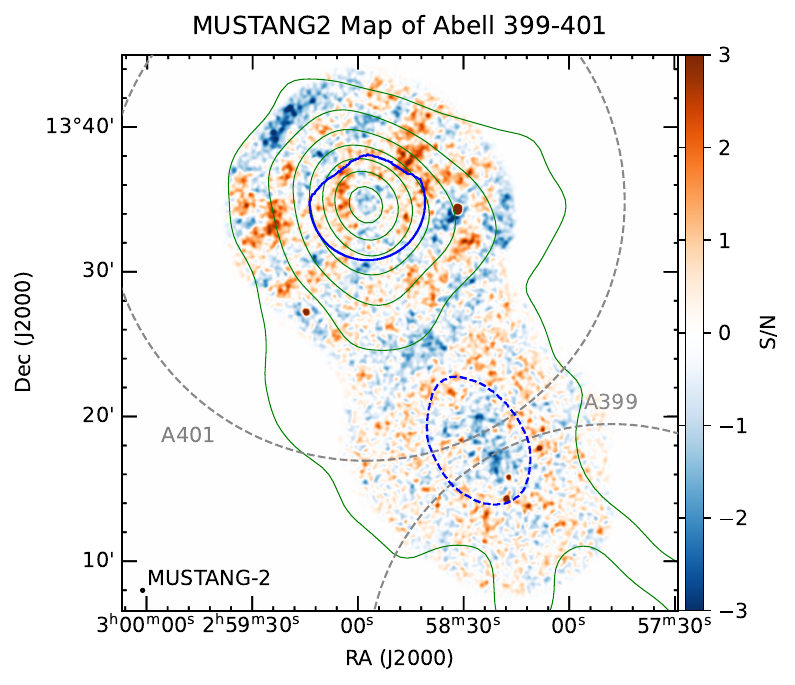}}
    {\includegraphics[clip, trim=0.0cm 0.32cm 0.2cm 0.7cm, height=6.9cm]{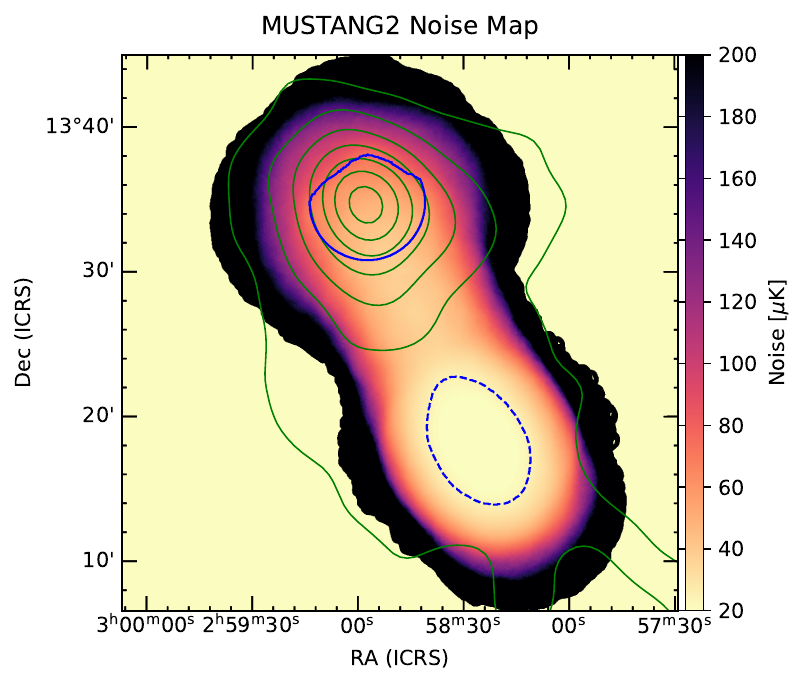}} 
    \caption{The images show (top) the MUSTANG-2 signal-to-noise (S/N) and (bottom) the MUSTANG-2 RMS (noise per beam) map, respectively.  The effective resolution (12.7\arcsec) of the smoothed map is depicted as a small circle in the lower left corner of the top panel. The regions used for pressure fluctuation analysis (Sec.~\ref{sec:features}) are shown with solid and dashed blue regions, corresponding to A401 and the bridge region, respectively.
    The contours and $R_{500\mathrm{c}}$ regions from the left panel of Fig.~\ref{fig:act_y_map} are overlaid (in green) for reference. We note that the region mapped by MUSTANG-2, and displayed here, was only a portion of that displayed in Fig. \ref{fig:act_y_map}, and A399 was not included in these MUSTANG-2 observations.}
    \label{fig:mustang2_maps}
\end{figure}

In order to construct our best estimate of the Compton-$y$ in the bridge region, we construct an Internal Linear Combination (ILC) of the available single-frequency maps in an 8 degree-wide region centred at right ascension 44.6 deg and declination 13.4 deg (approximately midway between A399 and A401), following closely the approach in \cite{madhavacheril/etal:2020}, to which we refer the reader for details. The procedure corresponds to calculating a linear combination of input maps, where the weights depend on an empirically measured covariance matrix and the response of each input map to the component of interest, i.e., Compton-$y$. The ILC uses the non-relativistic SZ formula; we correct for this approximation in our analysis (Sec.~\ref{ssec:total_mass}). The weights are designed to minimize the variance in the final map. Apart from the inclusion of data beyond what was collected by ACT up to 2015, there are three key differences with the analysis in \cite{madhavacheril/etal:2020}: 
\begin{enumerate}
    \item  Instead of starting from individual array maps, we start with the single-frequency \Planck + ACT coadds at 98, 150 and 220 GHz. These coadds are described in \cite{naess/etal:2020} for the DR5 release of maps including ACT data up to 2018; however, we use a more recent version that includes ACT data up to 2019. In order to control the contribution from the CIB and Galactic dust, we also include \Planck high-frequency channels at 353 and 545 GHz as done in \cite{madhavacheril/etal:2020}. 
    \item Since the coadding procedure in \cite{naess/etal:2020} has already optimally accounted for anisotropy of the ACT noise in 2D Fourier space, unlike \cite{madhavacheril/etal:2020}, we use a simplified noise model: specifically, to build the empirical covariance matrix, we simply calculate the empirical 1D auto- and cross-spectra between all input maps and bin these with bin widths of roughly $\Delta \ell = 250$ in order to reduce fluctuation-induced ILC bias  \citep[see, e.g.,][]{delabrouille/etal:2009}.  
    \item The bandpass information required for the Compton-$y$ response of the single-frequency \Planck + ACT co-adds is position-dependent and is constructed by linearly interpolating the weights provided with the \cite{naess/etal:2020} release (corresponding to the 2019 co-add) onto the centre of the patch on which our Compton-$y$ map is constructed.
\end{enumerate}
Fig. \ref{fig:act_y_map} shows the resulting map. Its edges, which are apodized during its construction, have been cropped to the $6.4\times6.4\,\si{\deg\squared}$ region shown in the left panel of the figure and used in this paper. It has a white noise level of $\ymapnoise$.\footnote{As estimated from the seven covariance fields: see Sec.~\ref{ssec:fit_procedure}.} The two clusters -- A401 in the north-east and A399 in the south-west -- are present with high significance and the elevated signal in the bridge between them is evident by eye. We characterize this quantitatively in Sec.~\ref{ssec:stats}.

Galactic dust and CIB are possible contaminants of the $y$-map. A visual inspection of the \Planck 857\,GHz map shows that the A399--401 region is relatively clean, though there could be some faint dust emission intruding upon the bridge region. While in principle the ILC should only contain SZ signal, one can add an additional constraint that requires a dust-like spectrum to have no response in the sum of the ILC weights. Thus, as a check we have also created a $y$-map with a dust spectrum deprojected in this manner, modelling it as a modified blackbody with a temperature of $24\,\si{\kelvin}$ and a spectral index of 1.2. See \citealt{madhavacheril/etal:2020} for details about this process. Note that there is uncertainty and spatial variation in the spectrum of Galactic dust and CIB, so the deprojection is a test for whether there are significant, dust-like residuals in the map, rather than the removal of a specific physical component. The resulting map is noisier ($\ymapdeprojnoise$ vs. $\ymapnoise$) and has a lower effective resolution ($2.2\arcmin$ vs. $1.65\arcmin$). We use the non-dust-deprojected map for our main analysis, but check that it is not significantly contaminated by running the same fits on this dust-deprojected map (Sec.~\ref{ssec:results}).

\subsection{MUSTANG-2 Data}
\label{ssec:mustang_data}

MUSTANG-2 is a 215-element array of feedhorn-coupled TES bolometers \citep{dicker/etal:2014}. Observing at 90 GHz on the 100-meter Green Bank Telescope (GBT), MUSTANG-2 achieves a resolution of 9\arcsec\ and has an instantaneous field of view (FOV) of 4.\!\arcmin25. 
It observes with on-the-fly mapping, typically using a Lissajous daisy scan pattern \citep[e.g.][]{dicker/etal:2014,romero/etal:2020}. 

We observed the filamentary region between A399 and A401, as well as A401 itself, between October 2019 and January 2020 (project code AGBT\_19B\_095), using the Lissajous daisy scan with a variety of scan radii (2.5, 3.0, 3.5, 5.0 and 6.0\,arcmin). The primary target was the bridge region, with less time spent on A401 (see Fig. \ref{fig:mustang2_maps}). In total, 44 hours of integration time on the sky were obtained. Due to time limitations, observations did not include A399. A maximal depth of $2.9\,\si{\micro\kelvin}$-arcmin was achieved over ${\sim}4\,\si{\arcmin\squared}$, with $47\,\si{\arcmin\squared}$ being mapped to $4.6\,\si{\micro\kelvin}$-arcmin or better (Fig.~\ref{fig:mustang2_maps}).\footnote{These noise figures are for a 183\,arcsec$^2$ beam size.} The data were processed with the MIDAS pipeline, which recovers signals on scales of $9\arcsec < \theta \lesssim 180\arcsec$ \citep{romero/etal:2020}. The signal-to-noise map, smoothed by a $9\arcsec$ FWHM Gaussian kernel that results in a final $12.7\arcsec$ resolution, is also shown in Fig.~\ref{fig:mustang2_maps}.

In this paper, we analyze our MUSTANG-2 map separately from the ACT+\Planck data. Folding it into the full ILC analysis will be investigated in a future analysis (see Sec.~\ref{ssec:searching_for_shocks}, below), as further research is required in order to properly treat its different noise properties and resolution, particularly on the boundaries of its relatively small coverage area, when combining it with these other data. For the present, we use it on its own to perform an initial search for small-scale features in the filament's gas (Sec.~\ref{ssec:mustang_analysis}).

\section{Filament Model Fits}
\label{sec:results}

\subsection{Models for the A399--401 System}
\label{ssec:models}

To confirm the presence of the bridge and determine its properties, we fit a few different models to the ACT+\Planck $y$-map (c.f., Table~\ref{tab:fit_results}). In this section we describe the elements used in the fits described in Sec.~\ref{ssec:fit_procedure}. 

In all models, the two clusters are described using the isothermal $\beta$-profile \citep{cavaliere/fusco-femiano/1978}:
\begin{equation}
    P(r) = k_\textsc{b} T_\mathrm{e}
           \frac{n_0}{ \left[1 + \right(\frac{r}{r_\mathrm{c}}\left)^{2}\right]^{\frac{3}{2}\beta}}
    \label{eq:beta-profile}
\end{equation}
where $r$ is the distance from the cluster centre, $n_0$ is the central electron density in the cluster, $r_c$ is the core radius and $\beta$ the slope. The profile given by equation~(\ref{eq:beta-profile}) is spherically symmetric; to allow for asphericity, in some of our models we fit an elliptical beta model following the formalism of \citet{hughes/birkinshaw:1998}. In this case, the expression $r / r_c$ in equation~(\ref{eq:beta-profile}) is replaced by:
\begin{equation}
 \frac{r}{r_\mathrm{c}} \longrightarrow \frac{\sqrt{x^2 + (y/R)^2}}{r_\mathrm{c}},
\end{equation}
where $r_\mathrm{c}$ becomes the core radius associated with the major axis, $x$ is the coordinate of the major axis, $y$ that of the minor axis, and $R$ is the ratio of the minor axis to the major axis.\footnote{\citet{hughes/birkinshaw:1998} use $e \equiv 1 / R$, i.e., the ratio of the major to the minor, in their expression, but we use $R$, which is bounded between 0 and 1, for computational purposes.} Analytic formulae for projecting the $\beta$-profile onto the plane of the sky are provided by \citet{hughes/birkinshaw:1998}; for the elliptical case, the cluster's axis of revolution is taken to be at an angle $\theta$ in the plane of the sky rotating north from west and at an inclination $i$ with respect to the line of sight. The out-of-plane inclination, $i$, is degenerate with $R$ and the central amplitude, $A = k_{\mathrm{B}}T_{\mathrm{e}}n_0$, so we simply set $i = 90\degree$. The free parameters we fit for each cluster are: the right ascension and declination of the cluster centre, the amplitude $A$, $\beta$ and $r_\mathrm{c}$; for elliptical fits, $\theta$ and $R$ are also free parameters.

\begin{figure}
    \centering
    {\includegraphics[width=0.99\columnwidth, clip, trim=0.25cm 0.25cm 0.3cm 0.86cm]{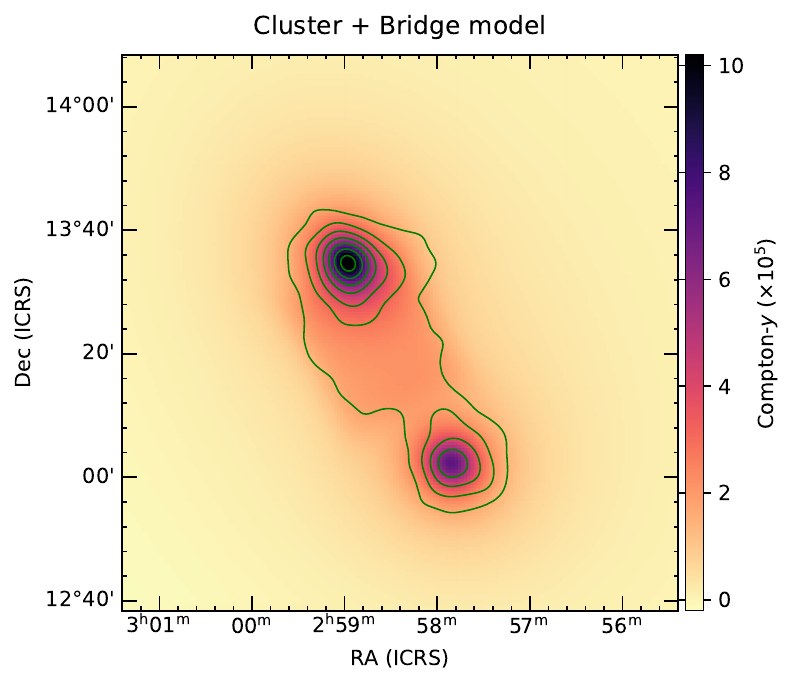}}
    \caption{The best-fit \modmesa model for the bridge and clusters, convolved to $1.65\arcmin$. The map is in units of Compton $y$, and shows the same region and contours as the right panel of Fig. \ref{fig:act_y_map}. Note that contours show levels in the data and not in the model.}
    \label{fig:2d_model}
\end{figure}

Unlike galaxy clusters, there is no established model for the bridge region. We try two \textit{ad hoc} models. In the first case, we simply fit a third beta model using the parameters described above, but with a fixed $\beta = 4/3$, appropriate for an isothermal cylinder in hydrostatic equilibrium \citep{ostriker:1964} and adopted by \citet{bonjean/etal:2018} in their cylindrical model. Note that this model is not intended to literally treat the bridge as a third cluster (c.f., the discussion at the end of Sec.~\ref{ssec:model_comparison}), but is simply an attempt to see if a profile with this power law can fit the extended signal in the bridge.\footnote{Note too that the $r_c$ obtained in the fit is extremely large compared to that of either of the actual clusters: see Table~\ref{tab:fit_results}.} In the second case, to capture the apparent flatness of the intercluster excess (see Fig.~\ref{fig:2d_model}) with minimal assumptions about its precise shape, we use the following `mesa' model:\footnote{We chose the name `mesa' because it has a relatively flat plateau that eventually falls off steeply: see Fig.~\ref{fig:slice_1D}.}
\begin{equation}
    g(l,w) = \frac{A_{\fil}}{1 + \left(\frac{l}{l_0}\right)^8 + \left(\frac{w}{w_0}\right)^8}.
	\label{eq:mesa}
\end{equation}
The coordinate system of equation~(\ref{eq:mesa}) is defined by an origin at the centre of the mesa, ($\alpha_{\fil}$, $\delta_{\fil}$), with $l$ being the axis parallel to the line joining the centres of the clusters (along the mesa length), and $w$ being the axis orthogonal to $l$ (along the width). The model has five free parameters: an amplitude $A_{\fil}$, the mesa centre ($\alpha_{\fil}$, $\delta_{\fil}$), and the characteristic size of the mesa length and width, $l_0$ and $w_0$, respectively.

Finally, in all cases we also fit a flat plane to model large-scale offsets and gradients in the background \citep{bonjean/etal:2018}:
\begin{equation}
    f(x, y) = a + bx + cy,
    \label{eq:background_offset}
\end{equation}
where $x$ and $y$ are map pixel coordinates. We adopt flat priors for all parameters, with the only constraints being that $2l_0$ not exceed the separation between A399 and A401, $0.4 < R < 1.0$ and $0 < \theta < -90$. For the starting values of the cluster centres in the Markov chain Monte Carlo (MCMC; see Sec.~\ref{ssec:fit_procedure}, below) we use the positions from Table~\ref{tab:basic_properties}.

\subsection{Fit Procedure}
\label{ssec:fit_procedure}

Using the parameters described in the previous section, we fit the central $128'\times128'$ region of the map (see Fig. \ref{fig:act_y_map}) with the following four models (c.f., Table~\ref{tab:fit_results}):
\begin{enumerate*}
    \item \modnobridge~-- elliptical $\beta$-profiles for the two clusters, with no model for the bridge;
    \item \modthreebeta~-- elliptical $\beta$-profile for the two clusters, plus an elliptical $\beta$-profile for the bridge region;
    \item \modcircmesa~-- azimuthally-symmetric $\beta$-profiles for the two clusters, with the mesa model for the bridge;
    \item \modmesa~-- elliptical $\beta$-profiles for the two clusters, with the mesa model for the bridge (shown in Fig. \ref{fig:2d_model}).
\end{enumerate*} 
The likelihood of a model given the data is:
\begin{equation}
    \mathcal{L} = \frac{1}{2 \pi |M|^{\frac{1}{2}}}
                  \exp\left(-\frac{1}{2}\mathbf{m}^{T}M^{-1}\mathbf{m}\right),
    \label{eq:likelihood}
\end{equation}
where $\mathbf{m}$ is the map of residuals -- i.e., the difference between the $y$-map and the model -- and $M$ is the covariance matrix of the map's noise. Before computing the residual, we convolve the model with a Gaussian kernel of $1.65'$ FWHM, which is the effective beam of our $y$-map. We estimate $M$ from the map itself using the seven fields around A399--401 that are indicated in Fig.~\ref{fig:act_y_map}; the field to the east of A399--401 is excluded due to obvious dust contamination. Although Compton-$y$ maps can be quite non-Gaussian due to the presence of tSZ signal \citep{madhavacheril/etal:2020}, these seven fields appear to be reasonably Gaussian---qualitatively similar to the fit residuals discussed below in Sec.~\ref{ssec:results} (see Fig.~\ref{fig:histograms}). We assess this by testing the normality of their $y$-values using the Kolmogorov--Smirnov (KS) test as well as the more stringent Anderson-Darling (AD) test \citep[e.g.,][]{stephens:1974}; we also try D'Agostino and Pearsons's (DP) omnibus test based on skewness and kurtosis \citep{dagostino/pearson:1973,dagostino/belanger/dagostino:1990}. All of the fields pass the KS test, and while four fields do not pass the other tests at any appreciable level, a visual inspection shows that they contain a few point-like fluctuations. When we mask the most prominent such sources in these fields (no more than three), the AD test improves notably and rises at least above the 2.5\% significance level; the DP omnibus test also improves, though in two cases falls slightly below the 1\% significance level. The origin of these fluctuations (which we stress are small) is not clear. The ILC procedure that produces our $y$-map begins with individual-frequency maps in which point sources $\gtrsim 5\sigma$ have already been removed \citep[see][]{madhavacheril/etal:2020}. Though one might hypothesise that fainter sources or imperfect subtraction causes these fluctuations, 4/10 of them do not coincide with the locations of any sources seen above $1\sigma$ in the individual-frequency ACT maps, so may represent true Compton-$y$ signals. Regardless, their presence does not affect our result: the covariance we compute (see below) is virtually identical whether we mask the fluctuations or the locations of known point sources or not. We discuss this further in Sec.~\ref{ssec:results}. In sum, while the fields are not completely free of non-Gaussian contamination, they should serve as a good proxy for the covariance $M$.

We compute equation~(\ref{eq:likelihood}) by using the convolution theorem. We square the fast Fourier transforms (FFT) of the noise fields described above, after apodizing the map edges by 29 pixels ($14.5'$) with a cosine function and correcting for the apodization weighting. The results are similar between the seven fields; we take their mean and smooth with an approximately Gaussian filter of $\sigma = 1.5$ pixels to obtain an estimate of the covariance $\hat{M}$ (where the hat indicates that a variable is in Fourier space). The precise choices of apodization width and smoothing scale do not significantly affect our best-fit values, but can impact the assessment of the model significances (Sec.~\ref{ssec:stats}) and are informed by simulations: see Appendix~\ref{appendix:fft_details} for details. Finally, we compute the FFT of the residuals, $\mathbf{\hat{m}}$, using the same apodization as the noise fields and calculate the sum of $\mathbf{\hat{m}}^*\mathbf{\hat{m}}/\hat{M}$. This yields the logarithm of the likelihood (modulo the constant factors in equation~\ref{eq:likelihood}).

The fits were performed by minimizing this log-likelihood with \textsc{emcee}  \citep{foreman-mackey/etal:2013}, a Python implementation of the affine-invariant MCMC algorithm designed by \citet{goodman/weare:2010}. We adopt a burn-in time of $10\tau$, where $\tau$ is the maximum of all parameters' autocorrelation lengths \citep[see][]{foreman-mackey/etal:2013}, and only include samples after this period in the calculation of our best-fit results. To verify that the measurement of $\tau$ was stable, we let the MCMC for the \modmesa model run for $> 40\tau$ additional iterations and found that it fluctuated by less than 1\% every 100 iterations. From this test we concluded that our MCMC fitting procedures are sufficiently converged.

\begin{figure*}
    \centering
    {\includegraphics[width=0.8\textwidth]{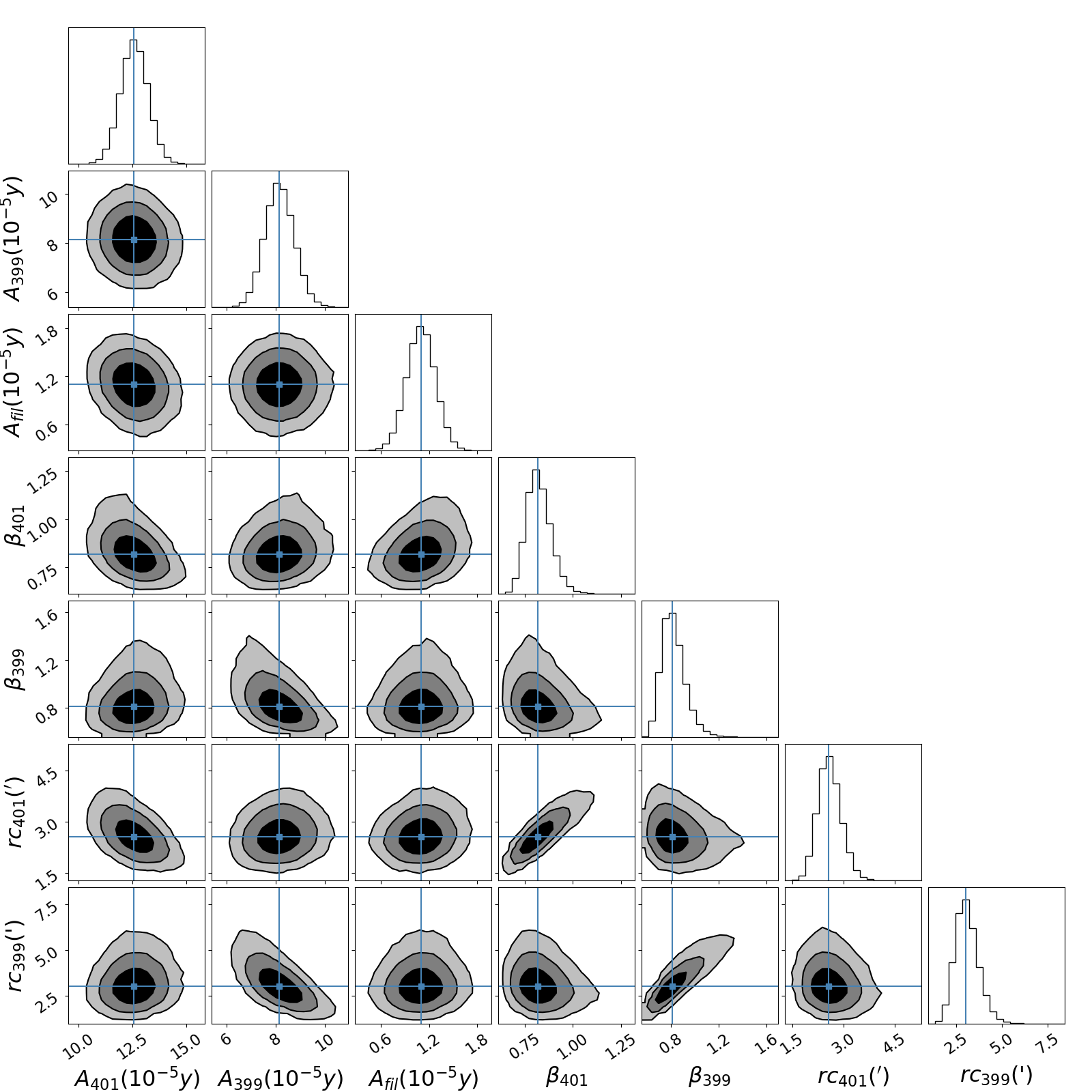}}
    \caption{Posterior distributions for a subset of parameters from the MCMC of the \modmesa model fit. Contours levels refer to the 68$^{\mathrm{th}}$, 95$^{\mathrm{th}}$ and 99.7$^{\mathrm{th}}$ percentiles. Best-fit parameters are indicated with blue lines. The expected degeneracy between $\beta$ and $r_c$, is clearly present. Some degeneracy does exist between other parameters (e.g., $r_c$ and $\beta$ and the background offset parameters of equation~\ref{eq:background_offset}, or the mesa/cluster amplitudes and coordinates), but those shown here are the most prominent.}
    \label{fig:contour_plot}
\end{figure*}

\subsection{Results}
\label{ssec:results}

\begin{table*}
    \caption{Best fit parameters for the models fit in this paper. The best values are the median of the posterior parameter distribution while the $1 \sigma$ errors refer to the $16^{\mathrm{th}}$ and $84^{\mathrm{th}}$ percentiles. Dashes indicate parameters that were not included in a given model. The background offset parameters $a$, $b$ and $c$ (equation~\ref{eq:background_offset}) are not shown below to prevent clutter; in all fits $|a| < 6\times10^{-6}$ and $|b|$, $|c| < 2\times10^{-8}$.}
    \label{tab:fit_results}
    \centering
    \begin{center}
        \begin{tabular}{cccccccc}
            \hline\hline
            \multicolumn{1}{c}{Model} & \multicolumn{7}{c}{A399} \\
             & $A$ ($10^{-5}y$) & $\alpha$($^{\circ}$) & $\delta$($^{\circ}$) & $\beta$ & $r_\mathrm{c}$ ($'$) & $\theta$ ($^{\circ}$) & $R$ \\
            \hline
            \modnobridge[1] &
                $8.2^{+0.6}_{-0.6}$ &  $ 44.474 \pm 0.003 $ & $ 13.030 \pm 0.003 $ & $0.74^{+0.08}_{-0.07}$ & $2.7^{+0.6}_{-0.5}$ & $-53^{+18}_{-18}$ & $0.87^{+0.07}_{-0.07}$ \\
            \modthreebeta[1] &
                $7.9^{+0.6}_{-0.6}$ & $ 44.473 \pm 0.004 $ & $ 13.030 \pm 0.003 $ & $0.84^{+0.13}_{-0.09}$ & $3.1^{+0.8}_{-0.6}$ & $-47^{+27}_{-26}$ & $0.92^{+0.05}_{-0.07}$ \\
            \modcircmesa[1] &
                $8.2^{+0.6}_{-0.6}$ & $ 44.473 \pm 0.004 $ & $ 13.030 \pm 0.003 $ & $0.80^{+0.10}_{-0.08}$ & $2.9^{+0.6}_{-0.5}$ & N/A & 1.0 \\
            \modmesa[1] &
                $8.1^{+0.6}_{-0.6}$ & $ 44.473 \pm 0.004 $ & $ 13.030 \pm 0.003 $ & $0.81^{+0.11}_{-0.08}$ & $3.0^{+0.7}_{-0.6} $ & $-47^{+27}_{-26}$ & $0.93^{+0.05}_{-0.07}$ \\
            \hline
        \end{tabular} 
    \end{center}
    
    \begin{center}
        \begin{tabular}{cccccccc}
            \hline\hline
            \multicolumn{1}{c}{Model} & \multicolumn{7}{c}{A401} \\
            & $A$ ($10^{-5}y$) & $\alpha$($^{\circ}$) & $\delta$($^{\circ}$) & $\beta$ & $r_\mathrm{c}$ ($'$) & $\theta$ ($^{\circ}$) & $R$ \\
            \hline
            \modnobridge[1] &
                $13.2^{+0.6}_{-0.6}$ & $ 44.750 \pm 0.002 $ &  $ 13.569 \pm 0.002 $ & $0.73^{+0.04}_{-0.04}$ & $2.4^{+0.3}_{-0.3}$ & $-61^{+6}_{-6}$ & $0.77^{+0.05}_{-0.05}$ \\
            \modthreebeta[1] &
                $11.8^{+0.8}_{-1.0}$ & $ 44.750 \pm 0.002 $ & $ 13.572 \pm 0.002 $ & $0.93^{+0.18}_{-0.10}$ & $2.9^{+0.5}_{-0.4}$ & $-56^{+10}_{-9}$ & $0.83^{+0.06}_{-0.05}$ \\
            \modcircmesa[1] &
                $12.7^{+0.6}_{-0.6}$ & $ 44.750 \pm 0.002 $ & $ 13.572 \pm 0.002 $ & $0.83^{+0.06}_{-0.05}$ & $2.4^{+0.3}_{-0.3}$ & N/A & 1.0 \\
            \modmesa[1] &
                $12.6^{+0.6}_{-0.6}$ &  $44.751 \pm 0.002$ & $13.572 \pm 0.002$ & $0.82^{+0.07}_{-0.06}$ & $2.6^{+0.4}_{-0.3}$ & $-57^{+9}_{-8}$ & $0.82^{+0.06}_{-0.05}$ \\
            \hline
        \end{tabular} 
    \end{center}
    
    \begin{center}
        \begin{tabular}{ccccccccc}
            \hline\hline
            \multicolumn{1}{c}{Model} & \multicolumn{8}{c}{Bridge} \\
            & $A_{\fil}$ ($10^{-5}y$) & $\alpha$($^{\circ}$) & $\delta$($^{\circ}$) & $l_0$ ($'$) & 
            $w_0$ ($'$) & $r_\mathrm{c}$ ($'$) & $\theta$ ($^{\circ}$) & $R$\\
            \hline
            \modnobridge[1] &
                --- & --- & --- & --- & --- & --- & --- & --- \\
            \modthreebeta[1] &
                $2.01^{+0.52}_{-0.41}$ & $44.69^{+0.03}_{-0.03}$ & $13.40^{+0.06}_{-0.05}$ & --- &
                --- & $16.0^{+1.7}_{-2.4}$ & fix & $0.80^{+0.13}_{-0.12}$ \\
            \modcircmesa[1] &
                $1.20^{+0.17}_{-0.17}$ & $44.67^{+0.01}_{-0.01}$ & $13.35^{+0.02}_{-0.03}$ & $11.7^{+1.4}_{-1.2}$ &
                $10.6^{+1.0}_{-0.9}$  & --- & fix & --- \\
            \modmesa[1] &
                $1.10^{+0.17}_{-0.18}$ & $44.68^{+0.02}_{-0.02}$ & $13.37^{+0.02}_{-0.03}$ & $12.3^{+1.7}_{-1.4}$ &
                $10.8^{+1.1}_{-1.0}$  & --- & fix & --- \\
            \hline
        \end{tabular} 
    \end{center}
\end{table*}

\begin{figure*}
    \centering
    {\includegraphics[height=4.25cm, clip, trim=00.0mm 3mm 23mm 6mm]{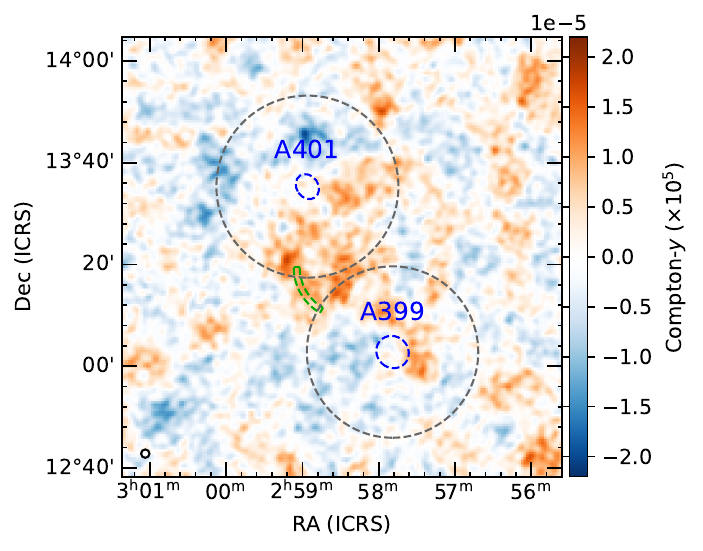}}
    {\includegraphics[height=4.25cm, clip, trim=19.5mm 3mm 23mm 6mm]{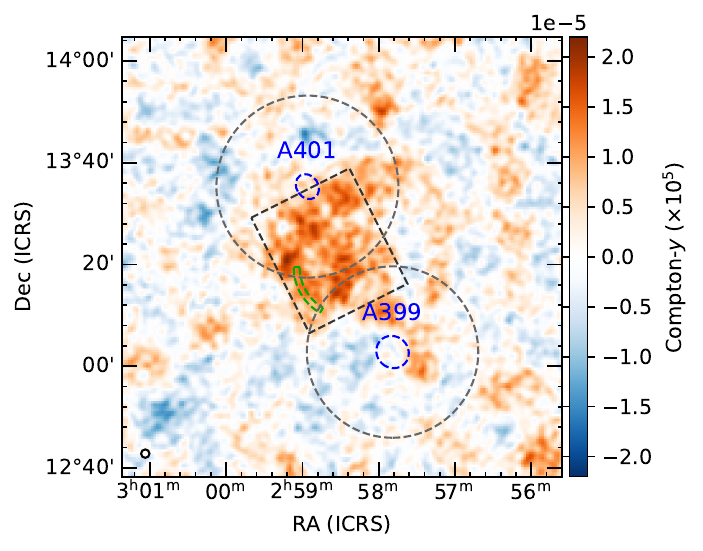}}
    {\includegraphics[height=4.25cm, clip, trim=19.5mm 3mm 23mm 6mm]{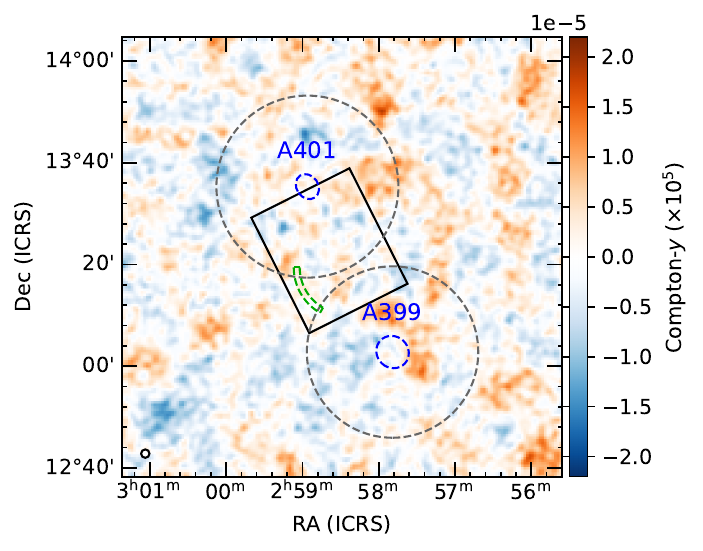}}
    {\includegraphics[height=4.25cm, clip, trim=19.5mm 3mm 03mm 6mm]{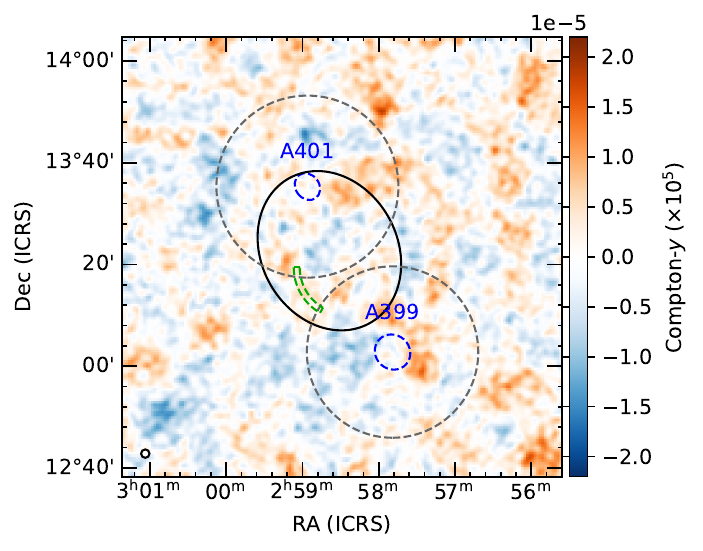}}    
    \caption{Residuals of the Compton-$y$ ACT+\Planck map after subtracting the best models. In all panels, the elliptical $\beta$-models for the clusters A401 and A399 have been subtracted.
    {\bf First:} The two cluster models in the \modnobridge model, which includes no bridge component, are subtracted.  A large residual at the location of the bridge is apparent.
    {\bf Second:} Only the models for the two clusters in the \modmesa model are subtracted; the mesa component (whose location is indicated with the dashed rectangle) is not subtracted.  A larger residual at the location of the bridge is apparent.
    {\bf Third:} The full \modmesa model (c.f., Fig.~\ref{fig:2d_model}) is subtracted, leaving mainly noise-like residuals.
    {\bf Fourth:} The \modthreebeta model is subtracted, leaving mainly noise-like residuals. A statistical analysis of all the fits is presented in Sec.~\ref{ssec:stats}.
    The dashed blue ellipses depict the position, angle, and major/minor axes of the core of the elliptical $\beta$-model fit to each cluster, while the dashed black circles centred at the same positions represent the value of $R_{500\mathrm{c}}$ for each cluster. The dashed/solid rectangles (middle panels) and solid oval (right panel) show the dimensions of the models for the bridge signal: the mesa ($2l_0 \times 2w_0$) and an extended $\beta$-profile ($r_c$), respectively. The central coordinates of the bridge component were free to vary, and the best-fit clearly prefers that the bridge be offset towards A401 and slightly to the east (see text).  The shock location identified in \citet{akamatsu/etal:2017} is depicted by the green dashed region. The very small, black circles in the bottom left corner of the panels indicate the $1.65'$ FWHM of the ACT beam.}
    \label{fig:2d_fit_results}
\end{figure*}

\begin{figure}
    \centering
    \includegraphics[width=\columnwidth]{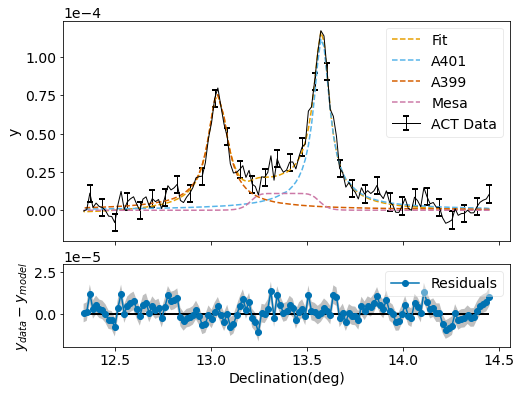}
    \caption{1D profiles along the axis connecting the centres of the two galaxy clusters together with the various components of the best fit \modmesa model and the sum of all model components (`Fit'). The residuals of the 1D profile are shown in the bottom panel. ACT error bars in the top panel are plotted for every four points to roughly indicate the space between independent points (the beam FWHM is 3.3 pixels). Error bars are the square root of the covariance matrix $M$ diagonal elements.
    Note that all our analysis is done with the 2D map; this figure is shown for illustration.}
    \label{fig:slice_1D}
\end{figure}

The MCMC best fit parameters are reported in Table~\ref{tab:fit_results}, while Fig.~\ref{fig:contour_plot} shows the posterior distributions for a selection of the parameters for the \modmesa model; as expected there are degeneracies between the cluster amplitudes $A$, their core radii $r_\mathrm{c}$, and the profile slope parameter $\beta$. Fig.~\ref{fig:2d_fit_results} compares map residuals with and without the components modelling the bridge subtracted. Fig.~\ref{fig:slice_1D} shows the profile of the $y$-map along the line connecting two galaxy cluster centres together with the profile of the best fit \modmesa model. Note that all our analyses are done on the 2D map (and that the positional parameters are free to vary jointly); the 1D profile is shown simply to provide a visualisation of the result. The goodness of these fits is analysed in below Sec.~\ref{ssec:stats}.

A striking feature of the bridge component of the fits is that it is not centred between the clusters but offset toward A401 (and also slightly to the east). We offer no physical interpretation of this result and defer investigation to future studies (see Sec.~\ref{ssec:searching_for_shocks}). Our focus at this time is on the overall interpretation of the presence of the excess gas, which we present in the following section.

We fit the same four models to the dust-deprojected map described in Sec.~\ref{ssec:act_data}. The field regions used for estimating the covariance in the dust-deprojected map are less Gaussian, and while a couple do pass the normality tests, the others are not readily ameliorated by masking prominent fluctuations, as these are more numerous. Nevertheless, we obtain fit results broadly consistent with those listed in Table~\ref{tab:fit_results}. On average the error bars are $\dustdeprojerrincrease$ larger with the dust deprojected map, as might be expected since it is slightly noisier and has poorer resolution. For all models, all parameters agree to within $1.5\sigma$ except for the right ascension of A399 (${\sim}2.4\sigma$ difference); due to the small uncertainty on this parameter, the difference is only ${\sim}1.4'$. The masses and other physical parameters for the best-fit model to our fiducial maps (Sec.~\ref{ssec:total_mass}--\ref{ssec:density_geometry}) agree with those from the dust-deprojected maps to within $1\sigma$; the results are included in Table~\ref{tab:fit_results_ddp}. We conclude that our results are not significantly affected by Galactic dust.

The residuals depicted in Fig.~\ref{fig:2d_fit_results} show qualitatively that the models that include a filamentary component fit the data better than the model that only fits A399 and A401 with elliptical $\beta$-models. In the next section we quantify this result, but let us first examine the goodness of the fits.

\begin{figure}
    \centering
    {\includegraphics[width=\columnwidth]{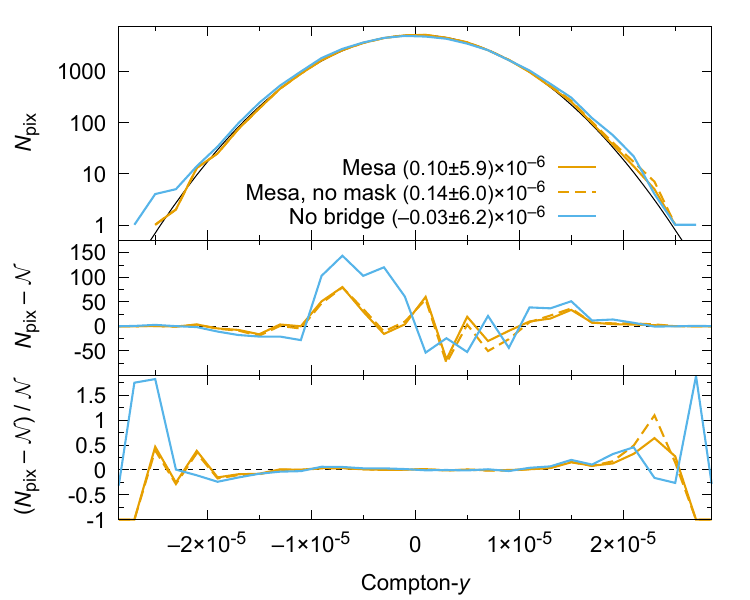}}
    \caption{Histograms of residual map values for the \modmesa and the \modnobridge models. The \modnobridge model exhibits a skewness as well as excess signal at the edges that cause it to fail all normality tests (see text and Table~\ref{tab:stats}). For the \modmesa model, we show the histogram both with and without masking the fluctuation NW of A401 (just outside $R_{500}$; see Fig.~\ref{fig:2d_fit_results}). Both pass the KS test, but masking the fluctuation significantly improves the AD and DP tests. \textit{Top:} Histograms for each residual map; the numbers in the caption indicate the mean and standard deviation of each distribution. The black line is the Gaussian with the mean and standard deviation from the \modmesa residual map. \textit{Middle:} Difference between the residuals and a Gaussian distribution; in this case, the mean and standard deviation of each individual map are used to compute the Gaussian. \textit{Bottom:} Same as the middle panel, except showing the fractional difference.}
    \label{fig:histograms}
\end{figure}

A common metric of fit quality is the probability-to-exceed (PTE) of the reduced $\chi^2$:
\begin{equation}
    \chi^2_r = \frac{\mathbf{m}^T M^{-1}\mathbf{m}}{K},
\end{equation}
where $\mathbf{m}$ and $M$ are defined in equation~(\ref{eq:likelihood}) and $K$ is the number of degrees of freedom (d.o.f.), often assumed to be the number of points, $N$, minus the number of fit parameters, $P$. However, assessing the true d.o.f. is difficult, especially for nonlinear models  \citep{andrae/schulze-hartung/melchior:2010}; furthermore, in our case the uncertainty in our covariance estimate and the necessity to apodize maps before the FFT leads to uncertainties in $\chi^2_r$ on the few-percent level (see Appendix~\ref{appendix:fft_details}). Still, we do expect $\chi^2_r \approx 1$ (using $K = N - P$) if our covariance estimate is robust. Correcting for the apodization weighting, we find $\chi^2_r = 1.01$ for all models. Another informative assessment of the goodness-of-fit is whether the residuals $\mathbf{m}$ are normally distributed \citep[see again][]{andrae/schulze-hartung/melchior:2010}. We run the same normality tests as we did on the noise regions (see Sec.~\ref{ssec:fit_procedure}), restricting ourselves to the interior $198\times198$ pixels that are not downweighted by the FFT apodization in the MCMC likelihoods.\footnote{There is a small, non-zero mean value to the residuals, smaller than the uncertainty of offset parameter $a$ (equation~\ref{eq:background_offset}). We find it necessary to subtract this residual mean before doing the KS test.} The results are shown in Table~\ref{tab:stats}. The residuals from the \modnobridge fit fail all the tests, including the most forgiving KS test (PTE = $1\times10^{-4}$). A skewness can be discerned in the histogram of these residuals in Fig.~\ref{fig:histograms}. The other models, which all include a fit for the bridge between the clusters, pass the KS test easily and pass the AD test at least the 1\% level, but this increases to better than 5\% level when the largest visible fluctuation in the map is masked (see Fig.~\ref{fig:act_y_map}, just NW of the $R_{500\mathrm{c}}$ boundary of A401).\footnote{If we include the tapered regions there are a few more fluctuations that must be masked to pass the AD test; they are, of course, quite far from the signal we want to fit.} As can be seen in Fig.~\ref{fig:histograms}, the change in the histogram is slight, demonstrating the sensitivity of these tests. The DP omnibus test appears to be the most stringent, and the low PTE values must be indicative of the low-level non-Gaussianity in our noise inferred earlier (Sec.~\ref{ssec:fit_procedure}). However, the overall picture is that the fits that include a bridge component are good, while the \modnobridge model does not fit the data well.

As a final check of the robustness of the fit against low-level contamination in the map, we ran the \modmesa model with the addition of four extra parameters representing the amplitudes of point sources that fall within the $3\sigma$ contour of our $y$-map (see Fig.~\ref{fig:act_y_map}).\footnote{We found the following NED matches: the sources near A401 are WISEA~J025911.46+133334.8 and PKS 0255+13 and those near A399 are WISEA~J025737.16+130049.3 and WISEA J025743.49+130941.8.} At none of these locations is there clear contamination, and when accounting for possible sources, resulting amplitudes are all $<1.6\sigma$, with the best fit values for the remaining parameters agreeing with those in Table~\ref{tab:fit_results} to within $0.2\sigma$. We also did a fit including these point sources in which we masked all point sources detected in our 90\,GHz map with SNR $> 1\sigma$ before estimating the covariance,\footnote{In the foregoing we used an updated version of the point source catalogue compared to that used in for the ILC but the two should be very similar.} and again found results consistent to better than $0.2\sigma$. We conclude that our result is not affected by point sources.

\section{Analysis of Filament Bulk Properties}
\label{sec:discussion}

\subsection{Evidence for Excess Gas in the Bridge}
\label{ssec:stats}

\begin{table*}
    \caption{Statistical information for model comparison. Shown are the number of parameters ($P$); the PTE for the normality of residuals according to the Kolmogorov--Smirnov (KS),  Anderson-Darling (AD) and D'Agostino and Pearsons's (DP) tests; the likelihood ratio ($W$) and $p$-value associated with the rejection of the null (listed as $1-p_{\mathrm{null}}$, i.e., the significance of the model in which the null model is nested) and its significance with reference to a normal distribution ($\sigma$); and the relative AIC ($\Delta$AIC) with its corresponding Akaike weight ($w_i$). Entries in the likelihood ratio columns with a dash (---) indicate that the statistic does not apply,  either because it is the null hypothesis (\modnobridge) or does not nest the null hypothesis (\modcircmesa). The top set of results is for our $y$-map that combines ACT and \Planck data; the lower set is for fits of the \Planck-only map (where the small-scale Fourier modes have been masked as described in the text). The normality tests are performed within the area where the FFT taper is unity, and, for the ACT+\Planck residuals, one large fluctuation has been masked. The AD test returns a metric that is assessed against a table with a finite number of PTE thresholds, which we indicate with the less-than or greater-than symbols.
    See text for more information and an interpretation of these statistics.}
    \label{tab:stats}
    \centering
    \begin{tabular}{ccccccccccc}
        \hline\hline
        Data & Model & $P$ & 
            \multicolumn{3}{c}{Normality PTE} &
            \multicolumn{3}{c}{Likelihood Ratio} &
            \multicolumn{2}{c}{AIC} \\
        & & & KS & AD & DP & $W$ & $1-p_{\mathrm{null}}$ & $\sigma$ & $\Delta_i$ & $w_i$ \\
        \hline
        ACT+\Planck & \modnobridge[1] & 
            17      &  1e-4     & < 0.01    & 1e-6  &
            ---     & ---                   & --- &
            33.4    & $4.3\times10^{-8}$ \\
        & \modthreebeta[1] &
            22      & 0.82      & > 0.15    & 0.03  &
            33.2    & $3.4\times10^{-6}$    & 4.6 &
            10.2    & 0.0047 \\
        & \modcircmesa[1] &
            18      & 0.65      & > 0.05    & 0.006 &
            ---           & ---             & --- &
            2.4     & 0.23 \\
        & \modmesa[1] &
            22      & 0.53      & > 0.05    & 0.009 &
            43.4    & $3.1\times10^{-8}$    & 5.5 &
            0       & 0.77 \\
        \hline
        \Planck only & \modnobridge[1] & 
            17      & 0.05      & > 0.15    & 0.20  &
            ---     & ---                   & ---   &
            13.52   & 0.0012 \\
        & \modmesa[1] &
            22      & 0.34      & > 0.10    & 0.001 &
            23.5    & $2.7\times10^{-4}$    & 3.6 &
            0       & 0.999 \\
        \hline
    \end{tabular}
\end{table*}

To quantify the significance of our detection of the bridge component, we test the null hypothesis -- i.e., the hypothesis that there is no filament -- with the likelihood-ratio statistic \citep[see, e.g.,][{\S}7]{held/bove:2014}:
\begin{equation}
    W = 2\log\frac{\max\mathcal{L}_2}{\max\mathcal{L}_1},
    \label{eq:likelihood_ratio}
\end{equation}
where $\max\mathcal{L}$ is the likelihood for the best-fit parameters (equation~\ref{eq:likelihood}) and the subscripts 1 and 2 denote two different models being compared. This statistic is only valid if Model 1 is `nested' in Model 2; that is, if Model 2 has the same parameters as Model 1 but extends it with additional parameters. On the hypothesis that Model 1 is the truth, then $W$ can be treated as having a $\chi^2$ distribution with d.o.f. equal to the number of extra parameters in Model 2. Because $W$ is a ratio, uncertainties in the likelihood introduced by the empirical covariance estimate tend to cancel out, making it an effective statistic for our purposes; we estimate that it has a precision of ${\sim}5\%$ (see Appendix~\ref{appendix:fft_details}). We use it to evaluate the `null hypothesis' of the \modnobridge model that is nested in both the \modthreebeta and the \modmesa models, each of which adds 5 parameters.\footnote{The \modcircmesa model is not nested in this null model, so we exclude it from the analysis here. However, it is nested in \modmesa. We compare these two models in the next section.} Table~\ref{tab:stats} shows the likelihood-ratio statistic for each of these scenarios, which have $W = \threebetaw$ and $W = \mesaw$, respectively; the respective $p$-values for the null hypothesis are $\threebetap$ and $\mesap$, or, alternatively, the bridge is preferred at about $\threebetapref$ when modelled with a third elliptical $\beta$-profile and about $\mesapref$ when modelled with the mesa of equation~(\ref{eq:mesa}). We thus make a significant detection of the bridge on the assumption that A399 and A401 can be modelled by elliptical $\beta$-profiles. Note that in the null hypothesis, the eccentricity, orientation and coordinates of each cluster profile are free parameters, so the bridge signal cannot be accounted for by elliptical clusters that overlap. Our result thus demonstrates that the observations are consistent with a filament of gas connecting the clusters.

To explore how much of our result is driven by the higher-resolution ACT data, we perform the same fits of the \modnobridge and \modmesa models on \Planck data only with the PR2 MILCA maps that have an $10\arcmin$ effective beam size \citep{Planck_PR2_15}, using the region around A399--401 resampled onto a flat projection with $3'$ pixels. We use an area of $2.5\times2.5\,\mathrm{deg}^2$, slightly larger than the area for our ACT+\Planck analysis (see Fig.~\ref{fig:act_y_map}), but choose an FFT apodization of $21'$ such that the area receiving full weight in the fits is close to the ACT+\Planck case. The covariance matrix was determined from five fields, not identical to those used for the ACT+\Planck map in order to avoid regions with significant non-Gaussian noise features, but still near to the A399-401 region. They are reasonably Gaussian according to the criteria used for our fiducial map (Sec.~\ref{ssec:fit_procedure}), though one field fails the AD and DP omnibus tests even with the few most obvious fluctuations masked; this notwithstanding, the resulting covariance estimate appears to be sound (see Appendix~\ref{appendix:fft_details}).
We do observe small-scale fluctuations in the map that may come from our flat-sky reprojection; however, in Fourier space these modes do not overlap with those pertaining to our fitting models, contributing a fraction of the covariance-weighted power of the models of order a few$\times10^{-5}$. We thus mask them and obtain $\chi^2_{\mathrm{r}} = 1.05$ (1.06) for the \modmesa (\modnobridge) model, as well as the statistics shown in the lower panel of Table~\ref{tab:stats}.
The residuals are reasonably normal; only the DP omnibus test fails for \modmesa, but is significantly improved by masking two fluctuations. One striking difference from our ACT+\Planck map is that the residuals of the \modnobridge model do not fail the normality tests. However, the likelihood-ratio test shows that the mesa model for the bridge is still preferred at $\mesaprefplanck$, a reduction of about $2\sigma$ compared to the result that includes ACT data.
Thus, if our analysis of the \Planck data alone provides evidence for the bridge, adding the ACT data provides a firm detection. Although the two previous analyses of \Planck data report significant detections of gas between the clusters \citep{planck/etal:2013,bonjean/etal:2018}, we note that neither of them tests the null hypothesis as we have here. \citet{bonjean/etal:2018} report ${\sim}8.5\sigma$ evidence for the gas by measuring the total signal-to-noise of the $y$-map in the region between the clusters that lies approximately outside of the clusters' $R_{500\mathrm{c}}$ boundaries.\footnote{Note that the $R_{500\mathrm{c}}$ values they use, which they obtain from a public database (http://szcluster-db.ias.u-psud.fr/) are ${\sim}15\%$ smaller than ours, at 1.2 and $1.3\,\si{\mega\parsec}$ for A399 and A401, respectively (c.f., Table~\ref{tab:phys_properties}).} Of course, this inevitably includes contributions from the clusters' outskirts in addition to any filamentary gas. However, they do separate the contributions in the their fit, which consists of the two cluster profiles plus a cylindrical bridge model, and make a $10.5\sigma$ measurement of the amplitude of the bridge's central pressure $P_0 = n_0 \, k_\textsc{b} T_{\mathrm{e}}$.\footnote{The earlier analysis of the \citet{planck/etal:2013} measured the bridge density to ${\sim}8\sigma$ in a joint fit with X-ray data. They do not explicitly calculate $P_0$, but it is readily determined from their $k_\textsc{b} T_{\mathrm{e}}$ and $n_0$ values, with errors added in quadrature.} In our analysis of the \Planck-only map, we find a best fit value of $A_{\fil} = \afilplanckonly$ for the \Planck-only map, a $\afilplanckonlysig$ measurement, quite a bit less significant. This may be because \citet{bonjean/etal:2018} use a model with fewer parameters. For instance, if we reduce the number of parameters in our \modmesa model by fixing the cluster and bridge locations (16 parameters compared to 10 in \citealt{bonjean/etal:2018}), the significance of our \Planck-only measurement of $A_{\fil}$ increases to $\afilplanckonlyfixpossig$. Alternatively, our use of a full noise covariance estimate could be responsible for increasing our uncertainties relative to theirs: \citet{bonjean/etal:2018} do not specify whether they use a full covariance treatment or simply assume white noise. If we set the off-diagonal elements of the covariance to zero in the fit to our full, 22 parameter model, our \Planck-only measurement of $A_{\fil}$ is at the $\afilplanckonlywhitenoisesig$ level, closer to their result.

\subsection{Comparison of Bridge Models}
\label{ssec:model_comparison}

In the previous subsection, we showed that the fits with a bridge component are good (via the normality of the residuals) and that a bridge component is required at the $\clustervsclusterbridge$ level (via the likelihood-ratio statistic). We now seek to compare the different models for the bridge itself. Since they are not all nested, the likelihood-ratio statistic cannot be used. Instead, we evaluate the Akaike Information Criterion \citep[AIC;][]{akaike:1974}:
\begin{equation}
    \AIC = 2K - 2\,\log(\max\mathcal{L}),
    \label{eq:aic}
\end{equation}
where $K = N - P$ is the number of data points (i.e., pixels in the map), $N$, minus the number of parameters, $P$.

Models with smaller AIC are better descriptions of the data and it is their relative differences, rather than absolute values, that are relevant. Let $\Delta_i = \AIC_i - \AIC_0 $ be the difference between the AIC of Model $i$ and the model with the smallest $\AIC$, which we call Model 0. Then the Akaike weights,
\begin{equation}
    w_i = \frac{\exp(-\Delta_i/2)}{\sum\exp(-\Delta_r/2)},
\end{equation}
where the sum in the denominator is over all models being compared, is a measure of the relative probability that Model $i$ is superior to the others \citep{burnham/anderson:2004}. More informally, $\Delta_i \gtrsim 4$ is considered to show `considerably less support' for model $i$; $\Delta_i \gtrsim 10$ shows `essentially no support' \citep{burnham/anderson:2004}. Table~\ref{tab:stats} lists $\Delta_i$ for our models. The model with the lowest AIC ($\AIC_0$) is \modmesa. As expected, the AIC indicates that the \modnobridge model is clearly disfavoured; furthermore, the \modthreebeta is a significantly worse fit than those that use our \textit{ad hoc} `mesa' model (equation~\ref{eq:mesa}) to describe the bridge region. We conclude that the filament is better represented by a relatively flat, rectangular model than by a model that peaks more strongly at the centre, like a $\beta$-profile (equation~\ref{eq:beta-profile}). The residuals for a $\beta$-model bridge in the rightmost panel of Fig.~\ref{fig:2d_fit_results} hint at this statistical preference: the more negative residuals at the centre of the bridge indicate an over-subtraction.\footnote{Another popular metric for model comparison is the Bayesian information criterion: BIC = $K\log(N) - 2\log(\max{\mathcal{L}})$, where $N$ is the number of samples \citep{schwarz:1978}. However, it does not function well on our data because the noise-dominated regions of our maps (i.e., far from A399--401) represent a large fraction of $N$. Since these regions are fit equally well by all our models, the penalty term $K\log{N}$ ends up inducing the BIC to prefer the model without a bridge (i.e., lower $K$) despite the obvious poorness of its fit. Indeed, \citet{burnham/anderson:2004} note that there are situations in which the BIC can `underfit' data.}

The result $\Delta_i = \circvsellipaicw$ for the \modcircmesa model shows that elliptical rather than circular $\beta$-profiles for A399 and A401 are statistically preferred, even when the extra parameters they add are penalized (see equation~\ref{eq:aic}). However, the preference is not significant ($w_i = 0.77$ vs. $w_i = 0.23$). Since the \modcircmesa model is nested in the \modmesa, we can also use the likelihood-ratio test to see the level at which the null hypothesis of no ellipticity in A399 or A401 is rejected. We find $W = \circvsellipw$, which has a $p$-value of $\circvsellipp$ ($\circvsellipsig$), indicating only mild support for cluster ellipticity.\footnote{Interestingly, the AIC statistics for the fits to the dust-deprojected map (see Sec.~\ref{ssec:results}) show less difference between the three models with a bridge: $w_i = 0.01$, 0.82, 0.17 for \modthreebeta, \modcircmesa and \modcircmesa, respectively. In this case there is less, but not significantly less, support for fitting with elliptical clusters compared with circular clusters. The likelihood-ratio test rejects circular clusters at only \circvsellipsigddp.}

The fits that included point source amplitudes (see Sec.~\ref{ssec:results}) do not improve the goodness of fit, being mildly disfavoured by the AIC test ($\Delta_i$ = 4.8; or $\Delta_i$ = 4.6 for the fits with sources masked for the covariance estimate).

We close this subsection by noting that one might propose that the excess signal in the bridge region comes from a third cluster at a different redshift from A399 and A401 superimposed between them by chance. However, the \Suzaku X-ray spectrum \citep{fujita/etal:2008,akamatsu/etal:2017} shows the characteristic iron line complex around 6.7~keV (rest frame), which constrains the gas in the bridge to be at the redshift of A399--401. The \citet{planck/etal:2013} also shows that such a third cluster would have to be at very high redshift (so as to have low X-ray brightness consistent with measurements) with a mass much higher than allowed by standard cosmology.  Our SZ measurement, which is roughly similar to \Planck's (see below), is consistent with this conclusion.

\subsection{Total Mass of the Bridge}
\label{ssec:total_mass}

We use the statistically-preferred \modmesa model to infer the total mass contained in the inter-cluster filament. The Compton parameter integrated over all angles, $Y = d_\mathrm{A}^2\int y \, \mathrm{d}\Omega$ where $d_\mathrm{A}$ is the angular diameter distance to the filament and $\Omega$ the solid angle in steradians, is related to the total mass by the following equation, assuming isothermality:
\begin{eqnarray}
    M_{\text{gas}} &=\,& \left[
        \frac{m_\mathrm{e} c^2\mu_\mathrm{e} m_\mathrm{u}}
             {\sigma_\textsc{t} k_\text{B} T_\text{e}}
        \right] Y \nonumber\\
    &=\,& \left[(1.1 \times 10^{18})
        \,\si{\msun\per\mega\parsec\squared}\right] Y.
    \label{eq:mass_gas2}
\end{eqnarray}
where on the second line, we have evaluated the expression with the atomic mass, $m_\mathrm{u}$, an electron molecular weight of $\mu_e = 1.155$ \citep[e.g.,][]{AndersGrevesse1989,adam2020}, and using $k_{\mathrm{B}}T_{\mathrm{e}} = 6.5\,\si{\kilo\electronvolt}$ \citep{akamatsu/etal:2017}. As noted above, equation~(\ref{eq:mass_gas2}) assumes that the gas is isothermal. \citet{sakelliou/ponman:2004} report X-ray temperatures in $5'$ and $7'$ chunks along the bridge axis, which exhibit a decline in temperature from ${\sim}9$ to ${\sim}6\,\si{\kilo\electronvolt}$ moving from A401 to A399, but the error bars are large and consistent with the $(6.5\pm0.4)\,\si{\kilo\electronvolt}$ measured in a $6'\times4'$ rectangle near the bridge centre by \citet{akamatsu/etal:2017}. The latter authors also measure the temperature in $2'$ increments perpendicular to the cluster axis, and find that it breaks about $10'$ from the centre, not far from the edge of our mesa (see the green shock location indicated in all panels of Fig.~\ref{fig:2d_fit_results}). On balance, while our assumption of isothermality is obviously a simplification, it is not necessarily a bad approximation.

We determine $Y$ by integrating equation~(\ref{eq:mesa}) with our best-fit values; to estimate the $1\sigma$ uncertainty we take the mean of the 16 and 84\% percentiles from the posterior distribution of $Y$ given by the MCMC chain. Because our $y$-map is made with the non-relativistic limit for the SZ effect (equation~\ref{eq:tsz}), this value will be underestimated at the high gas temperature in the bridge. At 98 and 150\,GHz, the frequencies that dominate the ILC at our scales (see Sec.~\ref{ssec:act_data}), the fractional size of the relativistic correction is the same to within 1\%, so we simply scale our measured $Y$ by this common correction. At $6.5\,\si{\kilo\electronvolt}$, this means multiplying our value $Y$ by 1.04; we obtained this value with the series expansions for the relativistic calculations in \citet{nozawa/itoh/kohyama:2005}. 

We find $Y = \Ympc$ which translates, via equation~(\ref{eq:mass_gas2}), to a gas mass $M_{\text{gas}} = \bridgegasmass$. (This value and other physical properties derived below are summarized in Table~\ref{tab:phys_properties}.)  Our result can be compared to the value of $\suzakugasmass$ from \citet{akamatsu/etal:2017}, which they obtained by combining their \Suzaku X-ray measurement with \Planck SZ results. Note that this and other values from \citet{akamatsu/etal:2017} have been corrected for some numerical errors---see Appendix~\ref{appendix:suzaku_correction} for details. Their result is for the whole bridge, whereas we have separated the filamentary component (modelled with the mesa) from the component due to the cluster outskirts. We would therefore expect them to report a value at least ${\sim}2\times$ greater (see Fig.~\ref{fig:slice_1D}). However, they assume a bridge length of only $1\,\si{\mega\parsec}$ when calculating the mass; if we scale their mass to match our length of $l_{\fil} = \mesalengthmpcapprox$ (see below, Sec.~\ref{ssec:density_geometry}), this tension is  resolved. Finally, we can estimate the total mass of the filament under the assumption that the ratio of baryons to total mass is the cosmic value of $16\%$ \citep{aiola/etal:2020}, obtaining $M_{\fil} = \bridgemass$.

To compare this filament mass with that of the whole A399--401 system, we determine the cluster masses by numerically integrating our best-fit $\beta$-profiles (equation~\ref{eq:beta-profile}) and then applying the scaling relation between $Y$ and $M_{500\mathrm{c}}$ provided by \citet{arnaud/etal:2010}.\footnote{This bootstrapping procedure is described in more detail in Appendix A of \citet{romero/etal:2020}. It consists in finding the intersection of $Y(<R)$ computed from the model and the value for $Y(<R_{500\mathrm{c}})$ versus $R_{500\mathrm{c}}$ when assuming the \citet{arnaud/etal:2010} scaling relation, but does not account for intrinsic scatter as was done in \citet{hilton/etal:2020}.  As a result, the error bars on $M_{500\mathrm{c}}$ reported here are smaller.} Here, $M_{500\mathrm{c}}$ is the mass contained within a radius $R_{500\mathrm{c}}$, inside of which the average density of the cluster is $500$ times the critical density of the Universe. We correct our $Y$ values for the relativistic SZ effect as we did for the bridge, using the temperatures measured from \textit{XMM--Newton} observations ($7.2\,\si{\kilo\electronvolt}$ for A399 and $8.5\,\si{\kilo\electronvolt}$ for A401; \citealt{sakelliou/ponman:2004}). Following \citet{hilton/etal:2020}, we also correct the resulting mass for the richness-based weak-lensing calibration factor of $\langle M^{\mathrm{UPP}}_{500\mathrm{c}}\rangle / \langle M^{\lambda\mathrm{WL}}_{500\mathrm{c}}\rangle = 0.69\pm0.07$.\footnote{This value is slightly different than that provided by \citet{hilton/etal:2020}, viz., $0.71\pm0.07$, but the final catalogue uses $0.69\pm0.07$: see \url{https://lambda.gsfc.nasa.gov/product/act/actpol_dr5_szcluster_catalog_info.cfm}.} This yields $M_{500\mathrm{c},\mathrm{A399}} = \athreemassf$ and $M_{500\mathrm{c},\mathrm{A401}} = \afourmassf$, where the uncertainties are calculated from the MCMC as described for the bridge $Y$-value above. \citet{hilton/etal:2020} reported masses of $6.7^{+2.1}_{-1.7}\times10^{14}\,\msun$ and $9.7^{+2.9}_{-2.3}\times10^{14}\,\msun$, respectively.
Our results are consistent with these values, though we find larger masses for both clusters. Even though the inclusion of the mesa component acts to decrease the cluster mass measurement---the masses inferred from the \modnobridge fit parameters are $1.5\%$ and $9\%$ larger for A399 and A401, respectively\footnote{The mass reduction is due, of course, to the mesa absorbing some of the Compton-$y$ signal near the clusters; it is smaller for A399 since the mesa overlaps more with A401.}---the increased masses compared to \cite{hilton/etal:2020} are due to the outer slope of our best-fit $\beta$ model being shallower than that of the matched filter applied in that paper, which is based on the \citet{arnaud/etal:2010} universal pressure profile; hence we derive a higher mass at large radius. Since the cluster masses are not the focus of our paper, we do not further investigate this effect.
\citet{mason/myers:2000} report masses based on \textit{ROSAT} observations of $M_{500\mathrm{c},\mathrm{A399}}^{\mathit{ROSAT}} = (7.93\pm0.47)\times10^{14}\,\msun$ and $M_{500\mathrm{c},\mathrm{A401}}^{\mathit{ROSAT}} = 8.67^{+0.74}_{-0.47}\times10^{14}\,\msun$ (where we have inserted our fiducial Hubble constant into their result), in agreement with our results, though again our values are larger. \citet{sakelliou/ponman:2004} report masses from \textit{XMM--Newton} that are lower ($4.98\times10^{14}\,\msun$ and $6.13\times10^{14}\, \msun$, respectively) but as they do not quote uncertainties, the differences between our values and theirs cannot be properly evaluated.

To better estimate the total mass of the clusters, we convert $M_{500\mathrm{c}}$ to $M_{200\mathrm{m}}$ using the approach of \citet{hilton/etal:2020} which relies on relations provided by \citet{bhattacharya/etal:2013}; note that here $M_{200\mathrm{m}}$ is with respect to the mean density of the Universe rather than the critical density. We obtain $M_{200\mathrm{m},\mathrm{A399}} = \athreemasst$ and $M_{200\mathrm{m},\mathrm{A401}} = \afourmasst$. Comparing the cluster masses to $M_{\fil}$, we find that the bridge comprises $\bridgemasspercent$ of the total mass of the system. There are many possible systematics behind this value: our filament mass estimate assumes isothermality, relies on a single measurement of the gas temperature, assumes that the gas fraction is equivalent to the cosmic value; moreover, the cluster masses depend on $Y$--$M$ scaling relations, as well as their intrinsic ellipticities (see Sec.~\ref{ssec:density_geometry}, below). Finally, the division of which gas belongs to the clusters and which belongs to the filament is model-dependent. Nonetheless, our result shows that a significant portion of the mass of the system is associated with the filamentary structure, and thus the `missing baryons'. In the next subsection, we also provide evidence that the axis of the system is largely out of the plane of the sky, along the line of sight, so that we are seeing a foreshortened projection of a longer filament, further solidifying this conclusion. 

\citet{galarraga-espinosa/etal:2021} find in simulations that the majority of the SZ signal in filaments comes from gas bound in haloes (and therefore not WHIM, technically speaking). However, the Compton-$y$ signal in A399--401 is about two orders of magnitude larger than the average filament in their simulations, so their prediction does not necessarily obtain in our context. As an exercise, we apply SZ--mass scaling relations to estimate the amount of signal in our $y$-map that could originate from galaxy members. \citet{bonjean/etal:2018} found four galaxies with stellar masses of approximately $10^{11}\,\msun$ in the A399--401 bridge, and several more below this limit. 
First, we assume a value of $M_{\mathrm{stellar}} / M_{\mathrm{tot}} = 0.03$ \citep[consistent with, e.g.,][]{Kravtsov2018, Behroozi2019} and use the {\it Planck} $Y{-}M$ scaling relations (\citealt{Planck_2013_XX}; their relation is consistent with that of \citealt{arnaud/etal:2010}).  
\citet{Vikram2017} and \citet{Meinke2021} have shown that stacks of massive galaxies exhibit scalings approximately consistent with those for galaxy groups and clusters, which validates our application of the {\textit Planck} cluster scaling to these galaxies. We compute that each of the four galaxies would have an integrated $Y \sim 4\times10^{-9}\,\si{\mega\parsec\squared}$ at the system's redshift.
The value for $Y$ inferred from galaxies is thus extremely small compared to our measured $\Ympc$, and since $Y$ exhibits a steep scaling ($Y\propto M^{5/3}$ for self-similar scaling and $Y \propto M^{1.79 \pm 0.08}$ in \citealt{Planck_2013_XX}), the integrated SZ flux density drops ${\sim}50\times$ for each decade drop in galaxy mass. We conclude that barring a huge population of lower mass galaxies, the galaxy halo contribution to the A399--401 bridge mass is subdominant.
For this conservative estimate, we have ignored the  non-negligible dusty source contribution that should suppress the SZ signal at 98 and 150~GHz \cite[see, e.g.,][]{Meinke2021}.

We also consider the possibility that each of the massive galaxies considered above resides in a group that is roughly ten times larger.  In this case, each group would contribute $Y d_{\mathrm{A}}^2 \sim 2{-}3\times10^{-7}\,\si{\mega\parsec\squared}$,  still subdominant to the total value we attribute to the bridge.  Such a group would have a size of $R_{500}\sim 0.4{-}0.5\,\si{\mega\parsec}$, or roughly 5\,\arcmin, so its surface brightness would be too diffuse to be apparent in the ACT or MUSTANG-2 data.  However, the X-ray imaging presented in \citet{fujita/etal:1996} and \citet{akamatsu/etal:2017} suggests the emission is diffuse in nature, rather than localized to a small number (<\,10) of group-scale systems. A more detailed study of the X-ray structure, although of interest, is beyond the scope of our study which focuses on the SZ signal.

\subsection{Geometry, Density and Orientation of the Bridge}
\label{ssec:density_geometry}

We define the effective length and width, $l_{\fil}$ and $w_{\fil}$, of the mesa model such that they approximate it as a rectangular function, i.e.,
\begin{equation}
    l_{\fil} \times w_{\fil} = \iint_{-\infty}^{\infty}
        \frac{y\,\mathrm{d}l\,\mathrm{d}w}{A_{\mathrm{fil}}},
        \;\;\mathrm{with}\;\;l_{\fil} / w_{\fil} = l_0 / w_0,
\end{equation}
where $y$ is given by equation~(\ref{eq:mesa}). Performing the integral, one finds that $l_{\fil} = 2.085\,l_0$ and $w_{\fil} = 2.085\,w_0$. Using the best-fit values from Table~\ref{tab:fit_results}, we have $l_{\fil}\times w_{\fil} = \mesalengthmpc[0]\times\mesawidthmpc[0]\,\si{\mega\parsec\squared}$. This width can be compared to the rough estimate of ${\sim}2.6\,\si{\mega\parsec}$ from \citet{akamatsu/etal:2017}.

To find the central gas density, we need an estimate of the \textit{thickness} of the bridge along the line of sight, $r_{\fil}$ (i.e., $r_{\fil}$ is normal to $l_{\fil}$ and $w_{\fil}$, which are in the plane of the sky). We obtain this by combining our SZ amplitude with \Suzaku X-ray results from \citet{akamatsu/etal:2017}. They report a bridge temperature of $kT = (6.5\pm0.5)\,\si{\kilo\electronvolt}$ and an X-ray brightness that yields a Compton parameter of $y = (8.0\pm1.0)\times(r_{\fil}/\si{\mega\parsec})^{1/2}\times10^{-6}$. Note that due to an algebraic error, this is about half of the value reported in their paper: see Appendix~\ref{appendix:suzaku_correction}. Here, $y$ is the total Compton signal in the bridge, which the authors obtain from the \citet{planck/etal:2013} to solve for $r_{\fil}$ and a density $n_\mathrm{e}$. We can repeat this exercise using our value of $y_{\mathrm{tot}} = \bridgetotaly$ given by the best-fit value for the mesa plus the cluster outskirts at the centre of the mesa (which includes the relativistic correction). This position is located near the boundary of the $8'\times2'$ region used by \citet{akamatsu/etal:2017} for their X-ray measurement; using a $y$-value from a different location in the inter-cluster region would naturally alter our result (c.f., Fig.~\ref{fig:slice_1D}), but not by more than our quoted uncertainties.\footnote{E.g., if we took the midpoint between the clusters rather than the mesa centre, our gas density result would change by ${\sim}0.3\sigma$.} With our $y_{\mathrm{tot}}$ we obtain a thickness of $r_{\fil} = \mesaheightmpc$, which yields an electron density in the bridge of $n_\text{e} = \bridgedensity$. In our mesa model, about half is from the filament itself and half from the cluster outskirts (see Fig.~\ref{fig:slice_1D}). This translates to the filament having ${\sim}150$ times the mean baryon density of the Universe at its redshift.

This density is significantly lower than the value of $(4.3\pm0.7)\times10^{-4}\,\si{\per\centi\meter\cubed}$ calculated by \citet{bonjean/etal:2018} with \Planck data, which was similar to earlier determinations \citep{planck/etal:2013,akamatsu/etal:2017}.
However, in their modelling of the A399--401 system, these previous studies presume -- at least implicitly -- that the system lies entirely in the plane of the sky \citep[see also][]{akahori/yoshikawa:2008}.\footnote{This is not to say that they do not advert to a possible component of their separation along the line of sight (e.g., \citealt{akamatsu/etal:2017} mentions this explicitly), but rather that their bridge analyses do not account for it.} Using the \citet{akamatsu/etal:2017} measurement containing the algebraic error, \citet{bonjean/etal:2018} reported an $r_{\fil}$ that was about $4\times$ lower than the correct value, which happened to be about the same value as the filament width. This was taken as confirmation that their cylindrical model for the filament, with the cylinder axis (implicitly) in the plane of the sky, was correct. However, with the algebraic error corrected, $r_{\fil}$ is much larger; for their measurement of $y = (22.2\pm1.8)\times10^{-6}$, one finds $n_{\text{e}} = (1.1\pm0.3)\times10^{-4}\,\si{\per\centi\meter\cubed}$ and $r_{\fil} = (7.8\pm2.3)\,\si{\mega\parsec}$. These values are consistent with the values we found above of $n_{\text{e}} = \bridgedensity$ and $r_{\fil} = \mesaheightmpc$.

\begin{figure}
    \centering
    \includegraphics[width=1.0\columnwidth, clip, trim=0mm 13mm 10mm 24mm]{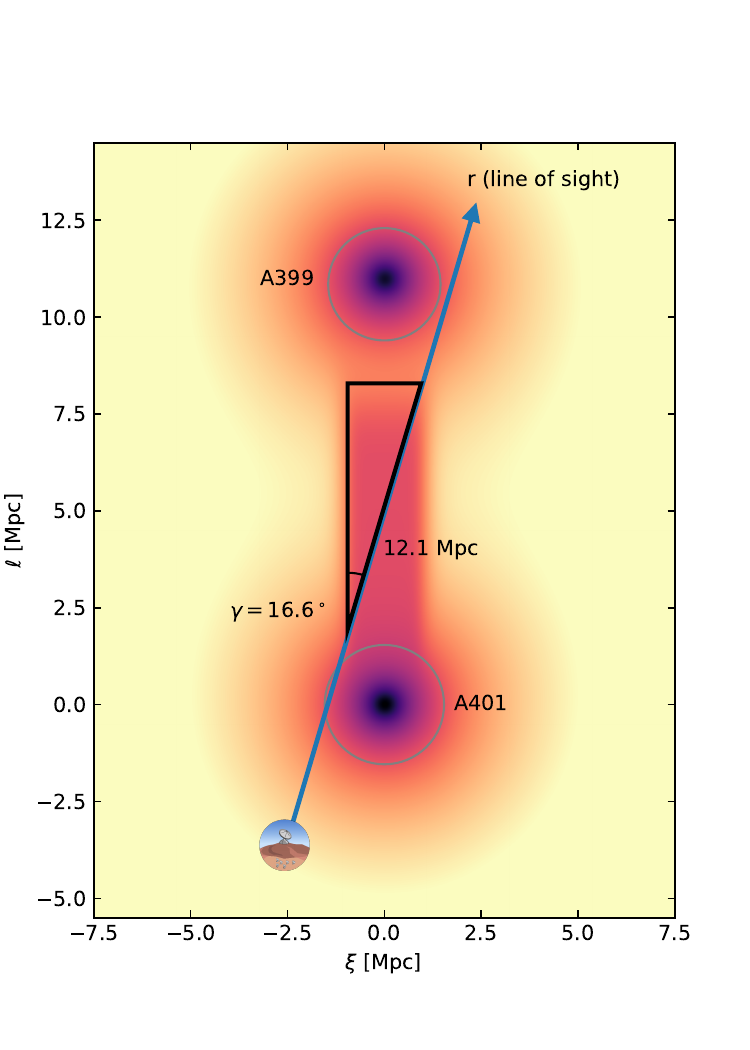}
    \caption{Toy model showing a slice of the gas pressure, displayed in a log scale, to illustrate the foreshortening of A399--401 in the plane of the sky. The line of sight of the observer is indicated in blue (i.e. A401, which is relatively redshifted compared to the median redshift, is closer to the observer assuming the difference in redshift is due to velocity rather than distance), and circles indicating $R_{500\mathrm{c}}$ are indicated as grey circles.}
    \label{fig:los_geometry}
\end{figure}

If we assume that the filament is roughly symmetric around the axis joining A399 and A401, our result suggests that the A399--401 system is significantly foreshortened. For instance, if the true thickness of the filament -- the dimension orthogonal to its width and length -- is equal to $w_{\fil}$ and all of $y_{\mathrm{c}}$ comes from a rectangular filament, then the angle between the line of sight and the A399--401 axis is $\gamma = \arcsin(w_{\fil} / r_{\fil}) \sim \toymodelnaifangle$. Fig.~\ref{fig:los_geometry} displays a toy model, based on the \modcircmesa geometry to avoid making assumptions about the inclinations $i$ of the cluster 3D ellipticities, and described in detail in Appendix~\ref{appendix:toy_model}). This toy model shows that the trigonometric approximation above is inadequate since the line of sight does graze the outskirts of the clusters. Taking the contribution from the cluster halos into account, this toy model yields an angle of $\gamma = \toymodelangle$. Given the separation between A399 and A401 in the plane of the sky of $d_{\mathrm{p}} = \clusterseparation$ (Table~\ref{tab:basic_properties}), this would imply a total separation of $\toymodelseparation$. We stress that this toy model is highly idealized. The basic result we wish to convey is chiefly qualitative: when we combine our SZ measurement of $y$ in the bridge with the \Suzaku X-ray data, we find that A399--401 axis has a significant component along the line of sight.

Given this orientation of the filament, we now explore the possible implications for the orientation of the clusters themselves, since our statistically-preferred \modmesa provides measurements of their ellipticities in the plane of the sky (the $R$ parameter in Table~\ref{tab:fit_results}). Cosmological simulations have found that cluster ellipticity has a statistical tendency to be co-aligned with local large scale filamentary structure \citep[e.g.,][]{kuchner/etal:2020}, and so we might expect that our clusters are approximately prolate ellipsoids with axes of revolution pointing along the filament. The true minor-to-major axis ratio $R_{\mathrm{t}}$ would then be related to the measured value in the plane of the sky, $R$, as \citep{fabricant/rybicki/gorenstein:1984}:
\begin{equation}
    R_{\mathrm{t}} = \frac{\sin\gamma}{\sqrt{1 - R^2\cos^2\gamma}}R \equiv F(R)R.
\end{equation}
The toy model presented above uses the \modcircmesa in order to remain agnostic about such cluster orientation (recall that this model is not significantly disfavoured compared to \modmesa). If we adapt the toy model such that it uses the \modmesa parameters and treats the clusters as prolate ellipsoids oriented along the filament axis, we obtain an angle consistent with that found above ($\gamma = \toymodelellipangle$) but with highly ellipsoidal clusters: $R_{399,\mathrm{t}} = \toymodelellipathreer$ and $R_{401,\mathrm{t}} = \toymodelellipafourr$. 
We can do a rough test of whether such ellipticities, which result from assuming that they are aligned with the filament axis, are reasonable by comparing to X-ray data. This approach works in an analogous way to our measurement of the bridge thickness above, since for a prolate ellipsoid the Compton-$y$ value is proportional to $\left[F(R)R\right]^{-1}$ \citep{fabricant/rybicki/gorenstein:1984,hughes/birkinshaw:1998}. Using the \textit{ROSAT} X-ray observations of A401 and A399, \cite{mason/myers:2000} modelled the clusters as spherical $\beta$-model systems and reported predicted SZ optical depths, $\tau_0 = \sigma_\textsc{t} \int n_{\mathrm{e}}(r)\mathrm{d}r$, which we can readily convert to a central Compton-$y$ value given the cluster temperatures they used.\footnote{They report values in terms of $\tau h^{1/2}$; we use our fiducial value of $h = H_0 / 70\,\si{\kilo\meter\per\second\per\mega\parsec} = 0.676$ in our calculation.} One obtains $y_\mathrm{A399} = (7.2\pm0.6)\times10^{-5}$ and $y_\mathrm{A401} = (13.3\pm1.2)\times10^{-5}$, which are in good agreement with our best fits (Table~\ref{tab:fit_results}). This suggests that $F(R)R \sim 1$ and that the clusters are unlikely to be significantly elongated or, in other words, that given the in-plane ellipticities we measure, it is unlikely that they are prolate with axes of revolution are aligned with the  filament.

Previous studies of intrinsic shapes of our clusters have been done through analyses of the precise shapes of their gas profiles and/or a combination of X-ray and SZ data. \citet{sereno/etal:2006} combined X-ray and SZ measurements and reported that A401 had a prolate geometry at an inclination of $(25\pm6)\,\mathrm{deg}$ and an intrinsic ellipticity of $R_{401,\mathrm{t}} = 0.46\pm0.47$, consistent with our elliptical toy model with the prolate ellipsoid axis aligned with the filament; however, they report an oblate shape for A399, which would be consistent with only mild elongation towards the line of sight, as suggested by the simple analysis we performed above. \citet{chakrabarty/defilippis/russell:2008} obtain a similar result for A399, but allow for triaxial geometries and find that A401 is close to oblate. These results would broadly indicate that A399 and A401 are not necessarily prolate along the filament. On the other hand, the relatively low values of $R$ we reported above from our toy model with prolate clusters are not unrealistic: \citet{sereno/etal:2006} find several examples with $R_{\mathrm{t}} \sim 0.3{-}0.4$ in their sample of 25 clusters.

From the numerical simulation side, \citet{lau/kravtsov/nagai:2009} and \citet{battaglia/etal:2012} present results compatible with the hypothesis that the clusters are not highly elongated, indicating that on sizes of $R_{500\mathrm{c}}$, cluster ellipticities are $R\sim0.8{-}0.9$ for gas density. However, simulations show that smaller values of $R$ are possible, particularly for unrelaxed clusters (\citealt{lau/kravtsov/nagai:2009}; as discussed further below, Sec.~\ref{ssec:mustang_analysis}, the relaxedness of our clusters is ambiguous). Though these authors do not explicitly address geometries of potentially interacting clusters, their results suggest that axis ratios close to unity are more prevalent; hence, while the possibility of our clusters being prolate and aligned with the filament is not ruled out by simulations, it appears less likely.

Since our main focus is the intercluster bridge, performing further analysis on the 3D ellipticity is beyond our scope and we simply conclude that our geometry is not unphysical. The only caveat is that if the clusters do have a large intrinsic ellipticity (\textit{contra} the arguments above), the mass estimates reported in Sec.~\ref{ssec:total_mass} could change significantly; this would alter the percentage of the total mass comprised by the bridge we quoted above (\bridgemasspercent), and would also have implications for the dynamics exercise we perform in the following subsection.

\subsection{Dynamics}

\begin{figure}
    \centering
    \includegraphics[width=1\columnwidth, clip, trim=2mm 5mm 2mm 2mm]{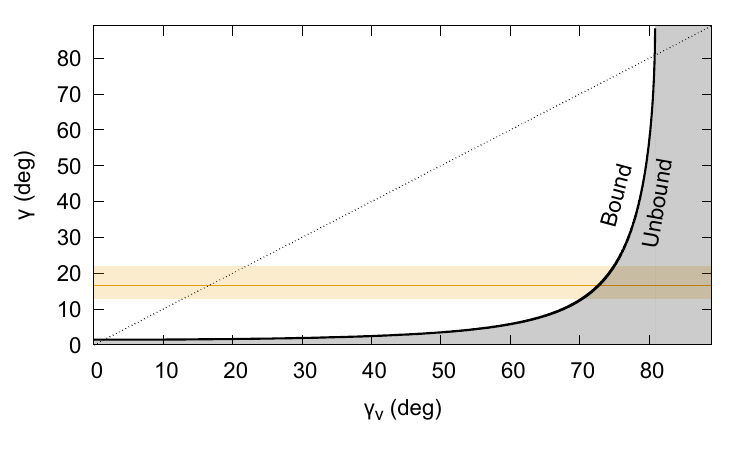}
    \caption{Conditions for whether A399 and A401 are gravitationally bound, depending upon the angle $\gamma$ between the axis joining the clusters and the line of sight, and the angle $\gamma_{\mathrm{v}}$ between the direction of the clusters' relative motion and the line of sight (equation~\ref{eq:gravitationally_bound}; c.f., fig.~3 in \citealt{hb1995}). The horizontal band shows the values for the cluster orientation given by our toy model, and the dashed line indicates alignment of the direction of relative motion with the axis joining the clusters. The graph shows that the system is bound unless the clusters have a large relative velocity in the plane of the sky ($\gamma_{\mathrm{v}} \gtrsim 70^{\circ}$).}
    \label{fig:bound}
\end{figure}

We close this section by commenting on the dynamics of the system. As noted above (Sec.~\ref{sec:introduction}), the X-ray and radio evidence indicates that the clusters are in a pre-merger state and moving together. This suggests that their relative motions have departed from the Hubble flow: after initially starting out close together and moving apart due to Hubble expansion, they eventually stalled under their mutual gravity and started falling towards each other. In this case, the difference between their redshifts (Table~\ref{tab:basic_properties}) can be treated as their velocity difference relative to the centre of mass of the system, projected along the line of sight; it evaluates to $v_\mathrm{r} = 520\,\si{\kilo\meter\per\second}$.\footnote{The correct formula, assuming $\Delta v \ll c$, is $\Delta v = c\Delta z / (1 + \bar{z})$, where $\Delta z$ is the difference of their redshifts, and $\bar{z}$ is their common redshift as they move away from us in the Hubble flow, which we take to be the mean of their redshifts \citep[see][]{davis/scrimgeour:2014}.} Note that A401 would be the more nearby cluster since it has the higher redshift, as depicted in Fig.~\ref{fig:los_geometry}. This dynamical picture only obtains if they are gravitationally bound, which can be tested by approximating A399--401 as a two-body system with the $M_{200\mathrm{m}}$ cluster masses reported in Sec.~\ref{ssec:total_mass}, and checking whether they satisfy the condition \citep{hb1995}:
\begin{equation}
    \left(\frac{v_{\mathrm{r}}}{\cos\gamma_{\mathrm{r}}}\right)^2 - 
    \frac{GM\sin\gamma}{d_\mathrm{p}} < 0,
    \label{eq:gravitationally_bound}
\end{equation}
where $\gamma_{\mathrm{r}}$ is the angle between the direction of their relative motion and the line of sight---for full generality, we do not assume in the equation that $\gamma_{\mathrm{r}} = \gamma$, i.e., that they are falling directly towards each other---$G$ is the gravitational constant and $M$ is the sum of the two cluster masses. Fig.~\ref{fig:bound} shows the parameter space for this binding for our values of $M$ and $d_{\mathrm{p}}$, and indicates that the system is unbound only if the largest component of the clusters' motion is perpendicular to the axis joining them (i.e., $|\gamma_{\mathrm{v}} - \gamma|$ is large). However, the presence of the filament between the clusters, which is ignored in the above formalism, makes this scenario hard to imagine, as large transverse cluster velocities would surely have severed or significantly disrupted the filament. We thus infer that the main component of the clusters' velocity is towards each other and that they are gravitationally bound. If we set $\gamma_{\mathrm{v}} = \gamma$ and take the orientation from our toy model, $\gamma = \toymodelangle$, the relative velocity is $v = \toymodelvpec$.\footnote{The uncertainty on $\gamma$ is not normally distributed, so to propagate the error to $v$ and the other values that follow in this paragraph, we calculate values using $\gamma + \mathrm{d}\gamma$ and $\gamma - \mathrm{d}\gamma$ and quote the resulting intervals as error bars. Since with our toy-model we cannot aim at high precision, this should suffice.} Given this velocity and the current separation, an analytic calculation of the Newtonian dynamics shows that the clusters' furthest separation was $\toymodelinitseparation$, $\toymodelinittime$ ago. Finally, using the same dynamics we can even provide a rough estimate age of the Universe: treating the clusters as starting out at the same initial location and moving apart, reaching the maximal separation just calculated and then moving back towards each other again, we find an age of $\toymodelageuniverse$: highly uncertain and model-dependent but consistent with the known age of the Universe.

\citet{yuan/etal:2005} performed a similar dynamical analysis using optical spectroscopic data \citep[see also][]{oegerle/hill:1994}, determining cluster masses via the galaxy velocity dispersions of $3.0\times10^{15}\,\msun$ and $3.1\times10^{15}\,\msun$ for A399 and A401, respectively (where here and in the following we have inserted our fiducial Hubble constant into their result). With a fixed age of the Universe ($13.9\,\si{\giga\year}$) and the assumption that the clusters began at the same place, they find $\gamma = (81.0\pm2.8)\,\mathrm{deg}$ or $(14.2^{+1.2}_{-1.3})\,\mathrm{deg}$ for the scenario where the clusters have already reached maximum separation and are approaching each other.\footnote{The authors also look at the case in which the clusters have already interacted and passed each other once, which they consider more likely based on the arguments of \citet{fabian/peres/white:1997}; as outlined in our Introduction, this is no longer considered plausible.} Given our results above in Sec.~\ref{ssec:density_geometry}, the first value would be ruled out, but there is good agreement between the second value with the angle we find with the toy model. For the case where the clusters have not reached maximum separation and are still outgoing, they find $\gamma = (8.4\pm0.6)\,\mathrm{deg}$; this scenario is, however, at odds with the proposal that the gas in the bridge is being heated by compression; see Sec.~\ref{sec:introduction}. The consistency between their analysis and ours indicates that our overall geometric picture is robust.

\begin{table}
    \caption{Derived physical properties of the A399--A401 system.}
    \label{tab:phys_properties}
    \centering
        \begin{tabular}{rlc}
            \hline\hline
            \multicolumn{2}{c}{Property} & Filament \\
            \hline
            Length & $l_{\fil}$ & \mesalengthmpc \\
            Width & $w_{\fil}$ & \mesawidthmpc \\
            Gas Mass & $M_{\mathrm{gas}}$ & \bridgegasmass \\
            Total Mass & $M_{\fil}$ & \bridgemass \\
            Thickness & $r_{\fil}$ & \mesaheightmpc \\
            Density & $n_\text{e}$ & \bridgedensity \\
            \hline
            \\
        \end{tabular}
        \\
        \begin{tabular}{ccc}
            \hline\hline
            Property & A399 & A401 \\
            \hline
            $R_{500\mathrm{c}}$ & \athreeradiusf & \afourradiusf \\
            $M_{500\mathrm{c}}$ & \athreemassf & \afourmassf \\
            $M_{200\mathrm{m}}$ & \athreemasst & \afourmasst \\
            \hline
        \end{tabular}
\end{table}

\section{Features Within the Filament}
\label{sec:features}

\subsection{Hints of Small-Scale Structure in MUSTANG-2 Data}
\label{ssec:mustang_analysis}

\begin{figure}
    \centering
 {\includegraphics[clip, trim=0.0cm 0.32cm 0.2cm 0.7cm, height=6.9cm]{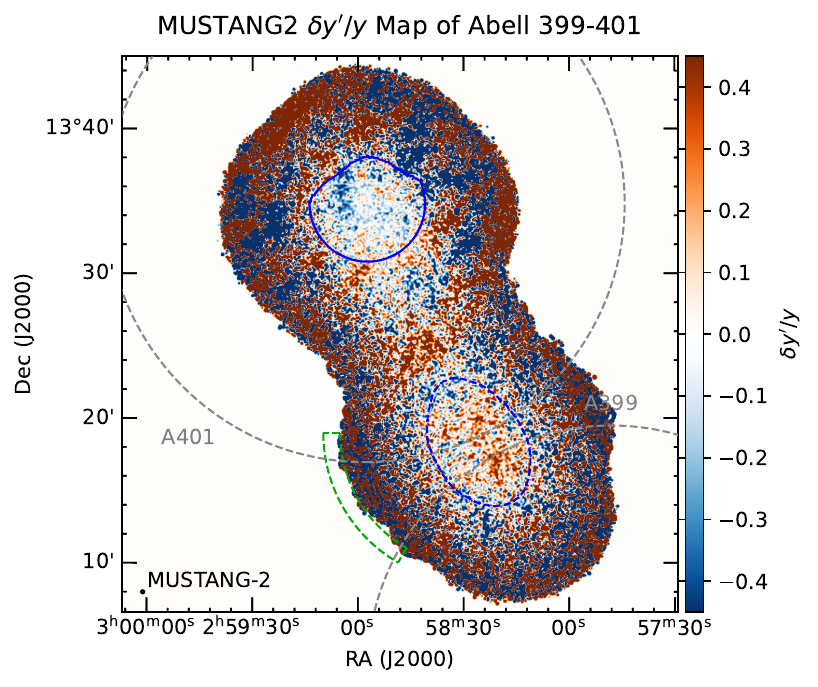}}
    \caption{ The $\delta y^{\prime} / y$ map of MUSTANG-2 data.  The effective resolution (12.7\arcsec) of the smoothed map is depicted as a small circle in the lower left corner of the top panel. The regions used for pressure fluctuation analysis are shown with solid and dashed blue regions, corresponding to A401 and the bridge region, respectively. The large absolute values near the edges of the map are reflect the higher noise where less time was spent observing.
    The location of the shock identified in \citet{akamatsu/etal:2017} is depicted by the green dashed region, and is just outside the coverage provided in the MUSTANG-2 observations reported. 
    The $R_{500\mathrm{c}}$ regions from the right panel of Fig.~\ref{fig:act_y_map} are overlaid for reference (grey dashed lines). We note that the region displayed is only a portion of that in Fig.~\ref{fig:act_y_map}, as the initial MUSTANG-2 observations did not map A399.
    }
    \label{fig:mustang2_dyy}
\end{figure}

Our 90\,GHz MUSTANG-2 temperature map only contains scales from $12.7\arcsec < \theta \lesssim 180\arcsec$ (see Sec.~\ref{ssec:mustang_data}), at which frequency and scales the signal is SZ-dominated. We convert it to a map of the Compton-$y$ parameter with equation~(\ref{eq:sz_spec}), and, to indicate that it only contains a restricted range of angular sensitivity, denote it a $y'$-map. Although this map is well-suited to finding small-scale features in the filament, we do not find clear evidence for any coherent substructure. To probe whether there are significant low-level fluctuations, we construct a map of $\delta y^{\prime} / \bar{y}$, where $\delta y^{\prime} = y^{\prime} - \bar{y}^{\prime}$, $\bar{y}$ is the best-fit \modmesa model (Sec.~\ref{ssec:fit_procedure}), and $\bar{y}^{\prime}$ is said model filtered through the MUSTANG-2 processing pipeline MIDAS. This map is shown in Fig.~\ref{fig:mustang2_dyy}. We examine the distribution of $\delta y^{\prime} / \bar{y}$ in both real space and log-normal space, i.e., $\log(1 + \delta y^{\prime} / \bar{y})$ and find that the former is more Gaussian. As the distribution for turbulence is expected to be log-normal \citep[e.g.,][]{khatri2016}, we infer that our $\delta y^{\prime} / \bar{y}$ is dominated by noise. Given the Gaussian distribution of $\delta y^{\prime} / \bar{y}$, the total variance represents the sum of the intrinsic variance and variance due to noise.

We thus investigate how large $\delta y^{\prime} / \bar{y}$ is compared to the expected map noise. To estimate the latter, we subtract the best-fit \modmesa model from the MUSTANG-2 timestreams. With the astronomical signal subtracted, we produce 100 noise realisations by reversing the order of the timestreams and then flipping the signs of (i.e., multiplying by $-1$) a random selection of half of the 304 scans before making the map. We then compute the average variance of $\delta y^{\prime} / \bar{y}$ from the noise realisations, and subtract that from the variance of $\delta y^{\prime} / \bar{y}$ in the real map. The result is our estimate of the amount of signal in $\delta y^{\prime} / \bar{y}$ fluctuations. In the region of the map with noise below $2.9\,\si{\micro\kelvin}$-arcmin (dashed contours in Fig.~\ref{fig:mustang2_dyy}) the rms of these intrinsic fluctuations is $0.078\pm0.015$. In the region of A401 where the noise is below $13.5\,\si{\micro\kelvin}$-arcmin  (solid contours in Fig.~\ref{fig:mustang2_dyy}), we find an rms of $0.052\pm0.022$.

In both cases, the inferred turbulence is rather low. A401 has been characterized both as non-relaxed \citep[e.g.,][]{bourdin/mazzotta:2008,parekh/etal:2015} and as relaxed \citep[e.g.,][]{vikhlinin/etal:2009,bonjean/etal:2018}, so its classification is debatable.\footnote{A399, on the other hand, is classified as non-relaxed by all these authors except \citet{bonjean/etal:2018}, who only mention its relaxed nature in a passing reference to \citet{sakelliou/ponman:2004}. The latter authors offer a more nuanced assessment.} The level of fluctuations we measure in A401 may suggest that there is no major merger activity, but would allow for minor ongoing mergers.
As for the fluctuations in the filament, there are few expectations yet set forth. Though turbulence (and thus fluctuations) should rise with cluster-centric radius, beyond the splashback radius gas motions may become more coherent (laminar). Alternatively, the geometry of the filament, which is potentially closely aligned with the line of sight (Sec.~\ref{ssec:density_geometry}), may conspire to wash out the fluctuations when projected on the plane of the sky.
 
\subsection{Searching for Shocks}
\label{ssec:searching_for_shocks}

Though we find no shocks in our MUSTANG-2 observations, they only cover a small portion of the bridge and could miss shocks further from its centre. A399--401 is believed to be in a pre-merger stage, with the clusters' motion possibly compressing and heating the WHIM in the filament between them (see Sec.~\ref{sec:introduction}). Simulations predict that this pressure along the axis joining the clusters should push the gas out in the perpendicular (or `equatorial') direction, creating shock fronts \citep{ha/ryu/kang:2018}. Such a front has been detected in X-ray measurements of another pre-merger system \citep{gu/etal:2019}. \citet{akamatsu/etal:2017} tentatively report an equatorial shock front in A399--401 to the south-east of the bridge, where they measure a steep drop-off in the gas temperature. We show its location in Fig.~\ref{fig:2d_fit_results} with the dashed green contour in all panels. Due to the weak evidence for a corresponding break in the X-ray brightness, \citet{gu/etal:2019} point out that the feature reported by \citet{akamatsu/etal:2017} could be due to `milder adiabatic compression'. Furthermore, with our finding that the system is largely out of the plane of the sky, the clusters' separation is perhaps larger than anticipated in the compression scenario. Still, in our map the feature is near the edge of the mesa, which could hint at the drop-off of a shock.

Clearly, better SZ observations are required in this region to confirm the presence of a shock. We plan to carry this out in a future study by adding recent, targeted observations of A399--401 with ACT, in combination with the MUSTANG-2 data. Such deeper, high-resolution maps might also discover shocks too faint to be observed in our current data.

These improvements could also elucidate the apparent offset of the intercluster gas toward A401 (see Figs.~\ref{fig:2d_fit_results} and \ref{fig:slice_1D}). If it is due to noise fluctuations artificially reducing the SZ signal near A399, the offset may be removed with better map sensitivity. On the other hand, the offset could be due to an overly-simplistic bridge model that does not adequately account for complex morphology in the cluster outskirts. In this case, the increased depth and resolution of new maps should allow us to probe this possibility. In a similar spirit, improved maps could also determine whether the arc- or handle-shaped feature that appears to join A401 and A399 on the west side of the system is real or is just a noise artifact. In our current maps this feature is slightly lower than $3\sigma$ above the white-noise level of the map (see Fig.~\ref{fig:act_y_map}, right panel).

\section{Summary \& Conclusions}
\label{sec:conclusion}

We have used a Compton-$y$ map of A399--401 created from ACT and \Planck data to confirm the presence of a filamentary structure between the two galaxy clusters. The high resolution and increased sensitivity of this map enable us to show that a model for the filament, fit jointly with $\beta$-profile models for the clusters, is required at ${\sim}\mesaprefapprox$. This represents an improvement in confidence over using a \Planck map alone (which yields a ${\sim}\mesaprefplanckapprox$ detection). Our fiducial model consists of two elliptical $\beta$ models, one for each cluster, and with a flat, `mesa-like' function, to which we assign no \textit{a priori} physical meaning (equation~\ref{eq:mesa}); there is significant statistical preference for this shape over the more rounded shape represented by a $\beta$-profile, as well as a small preference for the use of elliptical over symmetric cluster profiles. Of note is that even when the clusters are allowed to be elliptical, a bridge model is still required, making it implausible that the bridge signal is due solely to emission from the outer regions of clusters elongated along the axis separating them. The amplitude of the mesa, $y = \mesaamplitude$, is measured to ${\sim}6\sigma$ and contributes almost half of the total SZ signal in the centre of the mesa, with the rest contributed by the outskirts of A399 and A401.

A novel result of our investigation is the relatively low density of the filamentary region, $n_\mathrm{e} = \bridgedensity$, compared to previous measurements. We arrive at this figure by combining our SZ measurement with \Suzaku X-ray results \citep{akamatsu/etal:2017}, after correcting an algebraic error used in prior analyses. Our result implies that the thickness of the filament along the line of sight is significantly larger than its width in the plane of the sky: $\mesaheightmpcapprox$ compared to $\mesawidthmpcapprox$. We interpret this to mean that our view of the A399--401 system is considerably foreshortened, that is, the axis of the filament is mostly away from the plane of the sky. Constructing a simple toy simulation of the system based on our mesa model, we calculate that the axis is oriented at an angle of ${\sim}\toymodelangleapprox$ from the line of sight, implying a total separation of ${\sim}\toymodelseparationapprox$ between A399 and 401. These numbers come from a highly idealized scenario (a perfectly straight, cylindrical filament joining $\beta$-profile clusters), but it is reasonable to conclude that the picture of a mainly out-of-plane orientation is essentially correct, with the consequence that the excess gas in the intercluster region would be well beyond the $R_{500\mathrm{c}}$ radii of the clusters.  Finally, we show that the masses of the clusters and their relative velocities inferred from the redshifts are consistent with the system being gravitationally bound.

All of the above results combine to provide compelling evidence that we are seeing the gas in a long filament between two, premerger clusters. It provides further confirmation that a significant fraction of the Universe's baryons are located in the filamentary structure between galaxy clusters and, given the presence of radio emission from the bridge region \citep{govoni/etal:2019}, raises interesting questions about the nature of the magnetic fields in the cosmic web \citep[see, e.g.,][]{vazza/etal:2017}.

Finally, we have provided an initial analysis of high-resolution ($12.7\arcsec$) data of the bridge region from MUSTANG-2. We see no significant pressure substructure that might be induced by shocks, but do measure an excess signal above the noise which we attribute to low-level turbulence in the gas. This sets the stage for our planned future studies of A399--401. ACT has made additional observations of the system which we plan to use in a joint analysis with MUSTANG-2 data. While the focus of our current paper has been on characterising the bulk properties of A399--401, future work will explore the smaller-scale, lower-level dynamics of this system, which has proven to be a remarkable laboratory for studying cosmology and the astrophysics of the cosmic web.

\section*{Acknowledgements}

We thank Hiroki Akamatsu for his gracious and patient assistance in tracking down and correcting an algebraic error in \citet{akamatsu/etal:2017} (see Appendix~\ref{appendix:suzaku_correction}). We also thank Victor Bonjean and Nabila Aghanim for their help in understanding details from their work in \citet{bonjean/etal:2018}. We are grateful to the journal's scientific editor, Joop Schaye, and an anonymous referee for helpful suggestions.

This  work  was  supported  by  the  U.S.  National  Science Foundation  through  awards  AST-0408698,  AST-0965625,  and  AST-1440226  for  the  ACT  project,  as well as awards PHY-0355328,  PHY-0855887 and PHY-1214379.
Funding was also provided by Princeton University, the  University  of  Pennsylvania, and  a  Canada Foundation for Innovation (CFI) award to UBC.
ACT operates in the Parque Astron\'{o}mico Atacama in northern Chile under the auspices of the La Agencia Nacional de Investigaci\'{o}n y Desarrollo (ANID; formerly Comisi\'{o}n Nacional de Investigaci\'{o}n Cient\'{i}fica y Tecnol\'{o}gica de Chile, or CONICYT).
The development of multichroic detectors and lenses was supported by NASA grants NNX13AE56G and NNX14AB58G.
Detector research at NIST was supported by the NIST Innovations in Measurement Science program. Computations were performed on Cori at NERSC as part of the CMB Community allocation, on the Niagara supercomputer at the SciNet HPC Consortium, and on Feynman and Tiger at Princeton Research Computing, and on the hippo cluster at the University of KwaZulu-Natal. SciNet is funded by the CFI under the auspices of Compute Canada, the Government of Ontario, the Ontario Research Fund--Research Excellence, and the University of Toronto.
Colleagues at AstroNorte and RadioSky provide logistical support and keep operations in Chile running smoothly. We also thank the Mishrahi Fund and the Wilkinson Fund for their generous support of the project.

MUSTANG2 is supported by the NSF award number 1615604 and by the Mt.\ Cuba Astronomical Foundation. The Green Bank Observatory is a facility of the National Science Foundation operated under cooperative agreement by Associated Universities, Inc. GBT data were taken under the project ID AGBT\_19B\_095.

ADH is grateful for support from the Sutton Family Chair in Science, Christianity and Cultures.
CS acknowledges support from the Agencia Nacional de Investigaci\'on y Desarrollo (ANID) under FONDECYT grant no.\ 11191125.
EC acknowledges support from the STFC Ernest Rutherford Fellowship ST/M004856/2 and STFC Consolidated Grant ST/S00033X/1, and from the European Research Council (ERC) under the European Union’s Horizon 2020 research and innovation programme (Grant agreement No. 849169).  JLS acknowledges support from the Canada 150 Programme and an NSERC Discovery Grant.
JPH acknowledges funding for SZ cluster studies from NSF AAG number
AST-1615657.
KM acknowledges support from the National Research Foundation of South Africa. VV acknowledges support from INAF mainstream project ``Galaxy Clusters Science with LOFAR'' 1.05.01.86.05.
ZX is supported by the Gordon and Betty Moore Foundation.

This research has made use of the NASA/IPAC Extragalactic Database (NED), which is funded by the National Aeronautics and Space Administration and operated by the California Institute of Technology. 

Some of the results/plots in this paper have been derived/produced using the following software: 
\textsc{APLpy}, an open-source plotting package for Python \citep{software:aply/a,software:aply/b};
\textsc{Astropy},\footnote{http://www.astropy.org} a community-developed core Python package for Astronomy \citep{software:astropy/a,software:astropy/b};
\textsc{ds9} \citep{software:ds9};
\textsc{emcee} \citep{foreman-mackey/etal:2013};
\textsc{gnuplot},\footnote{http://www.gnuplot.info/}
\textsc{HEALPix}\footnote{http://healpix.sourceforge.net} \citep{software:healpix} and \textsc{healpy} \citep{software:healpy};
\textsc{Matplotlib} \citep{software:matplotlib};
\textsc{NumPy} \citep{software:numpy}; 
\textsc{pandas} \citep{software:pandas};
\textsc{pixell};\footnote{https://github.com/simonsobs/pixell} 
and \textsc{SciPy} \citep{software:scipy}.

\section*{Data Availability}

The ACT $y$-map used in this paper will be released on the NASA Legacy Archive Microwave Background Data Analysis (LAMBDA) website.\footnote{https://lambda.gsfc.nasa.gov/product/act/} The MUSTANG-2 map used in this paper will be released on LAMBDA and/or the Harvard Dataverse.\footnote{https://dataverse.harvard.edu/}

\bibliographystyle{mnras}
\bibliography{bridge.bib}

\appendix
\section{Some Details on the Noise Covariance Estimates}
\label{appendix:fft_details}

As described in Sec.~\ref{ssec:fit_procedure}, we calculate our likelihoods in Fourier space with particular choices for the edge-apodization and how we smooth our empirical covariance matrix. Altering these choices does not significantly change the fit results reported in Table~\ref{tab:fit_results} (e.g., even if the covariance matrix is not smoothed, the map fit area is shrunk and the apodization is reduced from 29 to 4 pixels). This is because while the overall amplitude of the likelihood can be sensitive to these parameters,  the MCMC only cares about the shape of the likelihood as a function of the fit parameters, regardless of the total amplitude. However, the amplitude of the best-fit likelihood \textit{can} affect the likelihood-ratio and AIC tests described in Secs.~\ref{ssec:stats}--\ref{ssec:model_comparison}. Here we describe how our results are impacted by this effect. Below, we refer to tests run both to our fiducial $y$-map (ACT+\Planck) as well as the \Planck PR2 MILCA-derived map used for the test in Sec.~\ref{ssec:stats} (\Planck-only).

Let us first consider the smoothing of the covariance. Since we are determining this matrix empirically by averaging the autocovariances of seven fields as described in Sec.~\ref{ssec:fit_procedure} (five in the case of \Planck-only; in all of the following, bracketed numbers represent the \Planck-only case), the resulting estimate has an inherent noise that has the tendency to bias $\chi^2 = \mathbf{m}^TM^{-1}\mathbf{m}$ high due to some values of $M$ scattering close to zero. To assess this bias, we take seven (five) fields of random, white noise with the same size and resolution as our $y$-maps, compute the empirical covariance as we do for the real maps, and then use them to determine $\chi^2$ from a map where the latter is known (i.e., a map of zeros with a single pixel of known amplitude). Repeating this process 128 times, we find that the bias on $\chi^2$ is ${\sim}15\%$ (${\sim}20\%$). It is reduced, however, by smoothing the resulting covariance estimate. For our smoothing kernel we use a 3-pixel wide moving average filter along each dimension of the array, repeated $N_{\mathrm{pass}}$ times---this approaches a Gaussian kernel as $N_{\mathrm{pass}}$ increases. For $N_{\mathrm{pass}}$ = 3 (approximately a Gaussian with $\sigma = 1.5$ pixels), the bias reduces to $1.0\%\pm0.3\%$ ($1.5\%\pm1.5\%$), where the uncertainty is the standard deviation from the 128 simulations. Turning to our data, when we compute the likelihood-ratio statistic $W$ (equation~9) using the best-fit residuals from our MCMC fit and our empirical covariance estimate with different $N_{\mathrm{pass}}$, we find that for $N_{\mathrm{pass}} \ge 3$, $W$ is stable to a few percent, while for $N_{\mathrm{pass}} < 3$ (less smoothing), $W$ becomes increasingly inflated.\footnote{Of course, if we smooth too much, $N_{\mathrm{pass}} \gg 1$, $W$ deviates from stability as the covariance estimate becomes poor.} We thus adopt $N_{\mathrm{pass}} = 3$.

Next we consider the width of the cosine apodization, $N_{\mathrm{apod}}$, that is applied to the edges of the maps before taking the FFT. All of our fitting models, which consist of $\beta$-profiles and the mesa profile (equations~\ref{eq:beta-profile} and \ref{eq:mesa}) have signal that continues to infinity. Although it drops well beyond the noise beyond the confines of the A399--401 system, if more area is included in the region used to compute likelihood, more of this model signal enters into the likelihood. The size of $N_{\mathrm{apod}}$, which effectively down-weights the signal at the map edge, thus has an impact on the likelihood ratio, $W$ (equation~\ref{eq:likelihood_ratio}). If, for instance, Model 2 has slightly more signal near the map edges than Model 1, then $W$ will decrease as the apodization grows, since more signal is being down-weighted.\footnote{This behaviour is not just a consequence of having models that continue to infinity. If one were to construct models with compact spatial support, by, for instance, introducing a cutoff to force them to zero at a certain distance, one would still need to make a choice about where that cutoff would be, and thus still determine the area over which the likelihood is determined.} Our choice of $N_{\mathrm{apod}} = 29$ pixels ($14.5'$, or 11\% of our map size on each side) seems to be reasonable. Varying $N_{\mathrm{apod}}$ between 19 and 59 only alters the likelihood ratios $W$ between our different models modestly, such that the significance of the bridge detection ($\mesapref$ for the \modmesa model; see Sec.~\ref{ssec:stats}) changes by ${\sim}{\pm}0.2\sigma$ in this range; below $N_{\mathrm{apod}} \sim 20$, $W$ rises more quickly as the map area increases, and soon the apodization becomes too narrow. (For \Planck-only, our taper is 7 pixels = $24'$, and we find that $W$ is very stable in the range of $N_{\mathrm{apod}}$ = 4--11.)

We do a similar test in which we take the residuals from the \modmesa, add a false, additional residual to the centre of the map to simulate a fit with no bridge component, and then compare that to the original residuals. By construction, this simulation has no difference in the residuals in the tapered region. In this case, $W$ remains fairly flat as $N_{\mathrm{apod}}$ is varied between 0 and 59, as expected, though the corresponding $\sigma$ does smoothly change by ${\sim}5\%$ in this range, perhaps due to decreased scale-mixing as the apodization improves. We also confirm that the smoothing scale of 3 is a reasonable choice by running this simulation with different smoothings. (For \Planck-only, we find a similar result.)

Summarising the foregoing, the statistics we quote in Secs.~\ref{ssec:stats}--\ref{ssec:model_comparison} are valid for a region of about $100\times100\,\si{\arcmin\squared}$ (i.e., the area within the apodized edges) centred on A399--401, and appear to be robust to within a few percent to modest changes in this effective area as well as to the choice of smoothing of the covariance estimate. This translates to uncertainties of ${\sim}0.2\sigma$ in the preference for the models with bridges over the \modnobridge, and uncertainties of ${\sim}1{-}2$ in the AIC parameter $\Delta_i$.

As a final check of the robustness of our method, we take the best-fit \modmesa model and fit its amplitude to the seven (five) fields used for the noise covariance:
\begin{equation}
    a_i = \frac{\mathbf{m}^T M^{-1} \mathbf{c}_i}{\mathbf{m}^T M^{-1} \mathbf{m}_i},
\end{equation}
where $\mathbf{m}$ is the map of the model, $M$ is the covariance estimate and $\mathbf{c}_i$ is the $i^{\mathrm{th}}$ field used in the covariance estimate. If $\mathbf{c}_i$ are true realisations of the covariance represented by $M$, then the mean of $a_i$ should be zero and their variance should be close to $\mathbf{m}^T M^{-1} \mathbf{m}$. Performing this test, we find variances that agree with the expectation to 5\% (12\%), which seems acceptable given the small sample size.

\section{Bridge Thickness, Density and Mass from \Suzaku Data}
\label{appendix:suzaku_correction}

In Sec.~4.1 of their paper, \citet{akamatsu/etal:2017} report a gas temperature of $k_{\mathrm{B}} T_{\rm e} = (6.5\pm0.5)\,\si{\kilo\electronvolt}$ in the intercluster region and an electron density:
\begin{equation}
    n_{\mathrm{e}} = (3.05\pm0.04)\times10^{-4}
        \left(\frac{r}{1\,\si{\mega\parsec}}\right)^{-1/2}
        \,\si{\per\centi\meter\cubed},
\end{equation}
where $r$ is the thickness along the line of sight as in our Sec.~\ref{ssec:density_geometry}, above. Assuming that $T_e$ and $n_\mathrm{e}$ are constant along the line of sight, one can write down (see equation~\ref{eq:tsz}, above):
\begin{equation}
    y = \sigma_{\textsc{t}} n_{\mathrm{e}}
        \frac{k_{\textsc{b}} T_{\mathrm{e}}}{m_{\mathrm{e}} c^2} r.
\end{equation}
The authors combine the above expressions and report a Compton parameter of:
\begin{equation}
    y_{\mathrm{orig}} = (14.5\pm1.8)\times10^{-6}
        \left(\frac{r}{1\,\si{\mega\parsec}}\right)^{1/2}
\end{equation}
However, two errors crept into this value, which we tracked down with the generous help of the lead author: (a) it has been multiplied by a stray factor of 2 and (b) a temperature of $6.0\,\si{\kilo\electronvolt}$ was used instead of $6.5\,\si{\kilo\electronvolt}$. When these are fixed, the correct value is (H. Akamatsu, private communication):
\begin{equation}
    y_{\mathrm{correct}} = (8.0\pm1.0)\times10^{-6}
        \left(\frac{r}{1\,\si{\mega\parsec}}\right)^{1/2}.
    \label{eq:y_akamatsu_corrected}
\end{equation}
This is the expression that we use for our results (Sec.~\ref{ssec:density_geometry}). Let us recalculate the quantities reported by the authors, which they determine with the $y$-value of $17\times10^{-6}$ taken from Fig.~2 of the \citet{planck/etal:2013}. Using equation~(\ref{eq:y_akamatsu_corrected}), their derived quantities change as follows:
\begin{eqnarray}
    r &=& 1.1\,\si{\mega\parsec} \rightarrow 4.6\,\si{\mega\parsec} \nonumber\\
    n_{\mathrm{e}} &=& 3.1\times10^{-4}\,\si{\centi\meter\cubed} \rightarrow 1.4\times10^{-4}\,\si{\centi\meter\cubed} \nonumber\\
    M_{\mathrm{gas}} &=& 1.3\times10^{13}\,\msun \rightarrow \suzakugasmass.
\end{eqnarray}
Note that $M_{\mathrm{gas}}$ has increased by more than the expected factor of ${\sim}2$; this is because \citet{akamatsu/etal:2017} used the mean molecular weight, rather than the electron molecular weight.\footnote{The authors of \citet{akamatsu/etal:2017} plan on adding an erratum with the above corrections to their paper.}

\section{Toy Model for Determining Geometry}
\label{appendix:toy_model}

Here we describe the toy model we use to estimate the angle between the A399--401 axis and the line-of-sight in Sec.~\ref{ssec:density_geometry}. It consists of a slice of the gas pressure down the axis joining the two clusters, as depicted in Fig.~\ref{fig:los_geometry}, that can be defined as follows:
\begin{equation}
    m(l, \xi) = \frac{\sigma_\textsc{t}}{m_{\mathrm{e}}c^2}P_{\mathrm{e}}(l, \xi),
    \label{eq:map_toy_model}
\end{equation}
where $l$ is, as defined in Sec.~\ref{ssec:fit_procedure}, the axis parallel to the line joining A399 and A401 and $\xi = r \cos\gamma'$, where $r$ is the line of sight and $\gamma'$ the angle between $r$ and $l$. The map is constructed at $w = 0$, i.e., a slice along the line joining A399 to A401 perpendicular to the plane of the sky, such that the integral along the coordinate $r$ is the Compton-$y$ parameter observed from our vantage point (equation~\ref{eq:tsz}):
\begin{equation}
    y(l) = \int m(l, \xi) dr.
    \label{eq:y_los}
\end{equation}
Computing the above expression yields a 1D profile such as that depicted in Fig.~\ref{fig:slice_1D}.

We create maps of equation~(\ref{eq:map_toy_model}) based on the mesa-model described in Sec.~\ref{ssec:fit_procedure}. Although the model that fits our data best uses elliptical $\beta$-profiles for A399 and A401, we do not know the inclination out of the plane of the sky, $i$, for either cluster, which would determine the ellipticity and orientation in the plane of the toy model. To avoid making any assumptions about this, we adopt our model that uses circular $\beta$-profiles, which we note provides almost as good a fit as the elliptical case ($\Delta_i = \circvsellipaicw$; see Table~\ref{tab:stats}).

For a given angle $\gamma'$, the toy model has the same geometry of the best-fit \modcircmesa model, except stretched by a factor of $1/\cos\gamma'$ and with the mesa horizontally centred on the A399--A401 axis. In more detail, we construct it as follows. We insert $\beta$-pressure profiles (equation~\ref{eq:beta-profile}) for each of A399 and A401 using the best-fit values (third row of Table~\ref{tab:fit_results}), separated by a distance of $l_{\mathrm{sep}} / \cos\gamma'$, where $l_{\mathrm{sep}}$ is the distance between the clusters' best-fit positions in the plane of the sky. We ensure the correct normalisation by numerically computing equation~(\ref{eq:y_los}) through the peak of each cluster individually and requiring that it equal their best-fit amplitudes $A$. Then, we include the bridge signal by adding the mesa model, equation~(\ref{eq:mesa}). We assume cylindrical symmetry, so that the characteristic size of the mesa along the $\xi$ axis is equal to $w_0$; along the $l$ axis we assign a size of $l_0 / \cos\gamma'$. We centre the mesa in the $\xi$ direction on the axis joining the clusters, and in the $l$ direction such that its position between A399 and A401, $l_{\mathrm{fil}}$, is proportionally the same as the best fit position after dividing by $\cos\gamma'$. Finally, for a given $\gamma'$, we set the amplitude of the mesa such that the total $y$ value measured in the centre of the mesa is the same as the best-fit value from the \modcircmesa  model ($y_{\mathrm{tot}} = \bridgecirctotalyapprox$).

We determine the best-fit $\gamma$ by creating maps of $m(l, \xi)$ between $\gamma' = 0$ and $90^{\circ}$ in $0.5^{\circ}$ increments. For each $\gamma'$ we numerically compute $y(l_{\mathrm{fil}})$ and from it derive the effective line-of-sight thickness $r'_{\mathrm{fil}} = y(l_{\mathrm{fil}}) / m(l_{\mathrm{fil}}, 0)$. The best fit occurs when $r'_{\mathrm{fil}}$ is closest to the value of $r_{\mathrm{fil}} = \mesaheightmpcapprox$ we derived by combining our $y$-value with the \Suzaku data in Sec.~\ref{ssec:density_geometry}. Our final determination of $\gamma$ is computed by linearly interpolating between the two models that are bisected by $r_{\mathrm{fil}}$.

We estimate an uncertainty by doing the above procedure at the upper and lower ends of the $1\sigma$ errors for $y_{\mathrm{tot}}$ (i.e., at $y_{\mathrm{tot}} - \mathrm{d}y_{\mathrm{tot}}$ and $y_{\mathrm{tot}} + \mathrm{d}y_{\mathrm{tot}}$), with $r_{\mathrm{fil}}$ also taken at its $\pm1\sigma$ bounds. The result, as used above in Sec.~\ref{ssec:density_geometry}, is $\gamma = \toymodelangle$. Note that $r_{\mathrm{fil}}$ was calculated in Sec.~\ref{ssec:density_geometry} using the $y_{\mathrm{tot}}$ from \modmesa model (and corrected for the relativistic SZ effect), while our toy model uses the \modcircmesa model. This seems to be the best approach since it uses our best measurement of $r_{\mathrm{fil}}$ and the \modcircmesa model has a bridge width almost identical to \modmesa. Still, as reported in Sec.~\ref{ssec:density_geometry}, when we use the latter model under the assumption that the clusters are prolate ellipses with axes aligned with the filament, we find $\gamma = \toymodelellipangle$, ${\sim}2^{\circ}$ lower, which illustrates the limitations of this toy-model. Since we claim no precision results based on $\gamma$, we do not attempt any further refinement.

\section{Dust-deprojected fit results}
\label{appendix:dust_depr}

Table~\ref{tab:fit_results_ddp} shows the best-fit values of our models using the dust deprojected $y$-map (see Sec.~\ref{ssec:results}). All parameters agree with the results from the fiducial $y$-map to $1.5\sigma$ except for the right ascension of A399 (${\sim}2.4\sigma)$), but this difference is smaller than our effective beam FWHM of $1.65'$. The agreement indicates that our results are not significantly contaminated by dust in the $y$-map.

\begin{table*}
    \caption{Best fit parameters for the models fit using the dust-deprojected map. The details on the entries are the same as for Table~\ref{tab:fit_results}.}
    \label{tab:fit_results_ddp}
    \centering
    \begin{center}
        \begin{tabular}{cccccccc}
            \hline\hline
            \multicolumn{1}{c}{Model} & \multicolumn{7}{c}{A399} \\
             & $A$ ($10^{-5}y$) & $\alpha$($^{\circ}$) & $\delta$($^{\circ}$) & $\beta$ & $r_\mathrm{c}$ ($'$) & $\theta$ ($^{\circ}$) & $R$ \\
            \hline\noalign{\smallskip}
            \modnobridge[1] &
                $7.4^{+0.9}_{-0.8}$ &  $ 44.451 \pm 0.005 $ &  $ 13.041 \pm 0.005 $ & $0.86^{+0.17}_{-0.12}$ & $4.0^{+1.3}_{-1.0}$ & $-57^{+14}_{-14}$ & $0.81^{+0.10}_{-0.09}$ \\\noalign{\smallskip}
            \modthreebeta[1] &
                $6.9^{+0.8}_{-0.6}$ & $ 44.449 \pm 0.005 $ & $ 13.038 \pm 0.005 $ & $1.11^{+0.30}_{-0.21}$ & $5.1^{+1.6}_{-1.3}$ & $-53^{+29}_{-22}$ & $0.90^{+0.06}_{-0.09}$ \\\noalign{\smallskip}
            \modcircmesa[1]&
                $7.2^{+0.8}_{-0.6}$ & $ 44.449 \pm 0.005 $ & $ 13.038 \pm 0.005 $ & $1.09^{+0.28}_{-0.19}$ & $4.8^{+1.5}_{-1.2} $ & $N/A$ & $1.0$ \\\noalign{\smallskip}
            \modmesa[1] &
                $7.2^{+0.8}_{-0.7}$ & $ 44.450 \pm 0.005 $ & $ 13.038 \pm 0.005 $ & $1.07^{+0.29}_{-0.19}$ & $5.0^{+1.5}_{-1.3}$ & $-53^{+32}_{-25}$ & $0.92^{+0.06}_{-0.08}$ \\\noalign{\smallskip}
            \hline
        \end{tabular} 
    \end{center}
    
    \begin{center}
        \begin{tabular}{cccccccc}
            \hline\hline
            \multicolumn{1}{c}{Model} & \multicolumn{7}{c}{A401} \\
            & $A$ ($10^{-5}y$) & $\alpha$($^{\circ}$) & $\delta$($^{\circ}$) & $\beta$ & $r_\mathrm{c}$ ($'$) & $\theta$ ($^{\circ}$) & $R$ \\
            \hline\noalign{\smallskip}
            \modnobridge[1] &
                $11.8^{+1.3}_{-1.1}$ & $ 44.737 \pm 0.003 $ & $ 13.574 \pm 0.004 $ & $0.76^{+0.08}_{-0.06}$ & $3.1^{+0.7}_{-0.7}$ & $-62^{+7}_{-7}$ & $0.75^{+0.06}_{-0.06}$ \\\noalign{\smallskip}
            \modthreebeta[1] &
                $10.3^{+1.3}_{-1.0}$ & $ 44.741 \pm 0.004 $ & $ 13.583 \pm 0.004 $ & $1.13^{+0.34}_{-0.22}$ & $3.9^{+1.1}_{-0.9}$ & $-47^{+16}_{-15}$ & $0.82^{+0.09}_{-0.08}$ \\\noalign{\smallskip}
            \modcircmesa[1] &
                $11.9^{+1.2}_{-1.0}$ &  $44.740 \pm 0.003$ & $13.580 \pm 0.004$ & $0.85^{+0.11}_{-0.08}$ & $2.8^{+0.7}_{-0.6}$ & $N/A$ & $1.0$ \\\noalign{\smallskip}
            \modmesa[1] &
                $11.9^{+1.3}_{-1.1}$ & $ 44.740 \pm 0.003 $ & $ 13.582 \pm 0.004 $ & $0.82^{+0.10}_{-0.08}$ & $2.9^{+0.8}_{-0.6}$ & $-51^{+14}_{-12}$ & $0.81^{+0.08}_{-0.08}$ \\\noalign{\smallskip}
            \hline
        \end{tabular} 
    \end{center}
    
    \begin{center}
        \begin{tabular}{ccccccccc}
            \hline\hline
            \multicolumn{1}{c}{Model} & \multicolumn{8}{c}{Bridge} \\
            & $A_{\fil}$ ($10^{-5}y$) & $\alpha$($^{\circ}$) & $\delta$($^{\circ}$) & $l_0$ ($'$) & 
            $w_0$ ($'$) & $r_\mathrm{c}$ ($'$) & $\theta$ ($^{\circ}$) & $R$\\
            \hline\noalign{\smallskip}
            \modnobridge[1] &
                --- & --- & --- & --- & --- & --- & --- & --- \\\noalign{\smallskip}
            \modthreebeta[1] &
                $2.41^{+0.58}_{-0.49}$ & $44.68^{+0.03}_{-0.03}$ & $13.41^{+0.05}_{-0.06}$ & --- &
                --- & $16.6^{+1.3}_{-2.1}$ & fix & $0.82^{+0.11}_{-0.11}$ \\\noalign{\smallskip}
            \modcircmesa[1] &
                $1.26^{+0.17}_{-0.18}$ & $44.66^{+0.02}_{-0.02}$ & $13.33^{+0.03}_{-0.02}$ & $12.3^{+1.3}_{-1.3}$ &
                $10.6^{+1.1}_{-1.1}$  & --- & fix & --- \\\noalign{\smallskip}
             \modmesa[1] &
                $1.16^{+0.19}_{-0.19}$ & $44.67^{+0.02}_{-0.02}$ & $13.35^{+0.03}_{-0.03}$ & $13.0^{+1.6}_{-1.6}$ &
                $10.7^{+1.5}_{-1.3}$  & --- & fix & --- \\\noalign{\smallskip}
            \hline
        \end{tabular} 
    \end{center}
\end{table*}

%
% Affiliations file automatically generated by the `make_author.py`.
% Better not to modify this by hand …
%
% Start affiliations
\section*{Affiliations}
\noindent\textit{%
$^{1}$David A. Dunlap Department of Astronomy \&Astrophysics, University of Toronto, 50 St. George St., Toronto ON M5S 3H4, Canada\\
$^{2}$Sapienza University of Rome, Physics Department, Piazzale Aldo Moro 5, 00185 Rome, Italy\\
$^{3}$Green Bank Observatory, 155 Observatory Road, Green Bank, WV 24944, USA\\
$^{4}$Department of Physics and Astronomy, University of Pennsylvania, 209 South 33rd Street, Philadelphia, PA 19104, USA\\
$^{5}$Centre for the Universe, Perimeter Institute, Waterloo ON N2L 2Y5, Canada\\
$^{6}$Department of Physics and Astronomy, University of Southern California, Los Angeles, CA 90007, USA\\
$^{7}$European Southern Observatory (ESO), Karl-Schwarzschild-Strasse 2, Garching 85748, Germany\\
$^{8}$Quantum Sensors Group, National Institute of Standards and Technology, Boulder, CO 80305, USA\\
$^{9}$Department of Astronomy, Cornell University, Ithaca, NY 14853, USA\\
$^{10}$Canadian Institute for Theoretical Astrophysics, University of Toronto, 60 St. George St., Toronto, ON M5S 3H4, Canada\\
$^{11}$School of Physics and Astronomy, Cardiff University, The Parade, Cardiff, CF24 3AA, UK\\
$^{12}$Joseph Henry Laboratories of Physics, Jadwin Hall, Princeton University, Princeton, NJ 08544, USA\\
$^{13}$Department of Astrophysical Sciences, Princeton University, Peyton Hall, Princeton, NJ 08544, USA\\
$^{14}$Instituto de Astrof\'isica and Centro de Astro-Ingenier\'ia, Facultad de F\'isica, Pontificia Universidad Cat\'olica de Chile, Av. Vicu\~na Mackenna 4860, 7820436, Macul, Santiago, Chile\\
$^{15}$Department of Physics, Cornell University, Ithaca, NY 14850, USA\\
$^{16}$INAF – Osservatorio Astronomico di Cagliari, Via della Scienza 5, I-09047 Selargius (CA), Italy\\
$^{17}$Department of Physics, Columbia University, New York, NY 10027, USA\\
$^{18}$Center for Computational Astrophysics, Flatiron Institute, New York, NY 10010, USA\\
$^{19}$Astrophysics Research Centre, University of KwaZulu-Natal, Westville Campus, Durban 4041, South Africa\\
$^{20}$Department of Physics and Astronomy, Rutgers, the State University of New Jersey, 136 Frelinghuysen Road, Piscataway,\\\hspace{0.85em} NJ 08854-8019, USA\\
$^{21}$Dunlap Institute of Astronomy \& Astrophysics, 50 St. George St., Toronto, ON M5S 3H4, Canada\\
$^{22}$National Radio Astronomy Observatory, 520 Edgemont Rd., Charlottesville, VA 22903, USA\\
$^{23}$Kavli Institute for Cosmological Physics, University of Chicago, Chicago, IL 60637, USA\\
$^{24}$Department of Astronomy and Astrophysics, University of Chicago, Chicago, IL 60637, USA\\
$^{25}$Department of Physics, University of Chicago, Chicago, IL 60637, USA\\
$^{26}$Enrico Fermi Institute, University of Chicago, Chicago, IL 60637, USA\\
$^{27}$School of Mathematics, Statistics \& Computer Science, University of KwaZulu-Natal, Westville Campus, Durban 4041, \\\hspace{0.85em} South Africa\\
$^{28}$Physics Department, Stanford University, Stanford,  CA 94305, USA\\
$^{29}$Kavli Institute for Particle Astrophysics and Cosmology, Stanford, CA 94305, USA\\
$^{30}$Department of Astronomy, University of Virginia, 530 McCormick Road, Charlottesville, VA 22904-4325, USA\\
$^{31}$Department of Physics, California Institute of Technology, Pasadena, CA 91125, USA\\
$^{32}$Department of Physics, McGill University, 3600 rue University, Montr\'{e}al QC H3A 2T8, Canada\\
$^{33}$McGill Space Institute, McGill University, 3550 rue University, Montr\'{e}al QC H3A 2A7, Canada\\
$^{34}$School of Chemistry \& Physics, University of KwaZulu-Natal, Westville Campus, Durban 4000, South Africa\\
$^{35}$Instituto de F\'isica, Pontificia Universidad Cat\'olica de Valpara\'iso, Casilla 4059, Valpara\'iso, Chile\\
$^{36}$School of Earth and Space Exploration, Arizona State University, Tempe, AZ 85287, USA\\
$^{37}$NASA Goddard Space Flight Center, Greenbelt, MD 20771, USA\\
$^{38}$MIT Kavli Institute, Massachusetts Institute of Technology, 77 Massachusetts Avenue, Cambridge, MA 02139, USA\\
}
%
% End affilations. End file.

\bsp	% typesetting comment
\label{lastpage}
\end{document}